\documentclass[10pt, pre, amsmath, twocolumn, showpacs, superscriptaddress]{revtex4-1}
\usepackage{amsmath}
\usepackage{amssymb}
\usepackage{mathtools}

\usepackage{color}
\definecolor{dgreen}{rgb}{0,0.7,0}

\usepackage{caption}
\usepackage{subcaption}

\newcommand{\Lx}{{L_x}}
\newcommand{\Ly}{{L_y}}
\newcommand{\Lyp}{{L_y'}}
\newcommand{\br}{\mathbf{r}}
\newcommand{\bR}{\mathbf{R}}
\newcommand{\bq}{\mathbf{q}}
\newcommand{\brT}{{\bR_T}}
\newcommand{\benu}{{\mathbf{e}_\nu}}
\newcommand{\bex}{{\mathbf{e}_x}}
\newcommand{\bey}{{\mathbf{e}_y}}
\newcommand{\bz}{\mathbf{0}}
\newcommand{\xT}{{X_T}}
\newcommand{\yT}{{Y_T}}
\newcommand{\dd}{\mathrm{d}}
\newcommand{\ee}{\mathrm{e}}
\newcommand{\ci}{\mathrm{i}}
\newcommand{\sgn}{\mathrm{sgn}}
\newcommand{\rhoz}{{\overline{\rho}}}
\newcommand{\hG}{\hat{G}}
\newcommand{\tG}{\tilde{G}}
\newcommand{\cA}{\mathcal{A}}
\newcommand{\hcA}{\hat{\mathcal{A}}}
\newcommand{\hcB}{\hat{\mathcal{B}}}
\newcommand{\pr}{P}
\newcommand{\kb}{{k_{\mathrm{B}}}}

\newcommand{\bn}{\mathbf{n}}
\newcommand{\ust}{{u^*}}
\newcommand{\vst}{{v^*}}
\newcommand{\must}{{\mu^*}}
\newcommand{\btau}{\mathbf{\tau}}
\newcommand{\Ap}{{A_+}}
\newcommand{\Am}{{A_-}}
\newcommand{\cZln}{\mathcal{Z}_{\lambda,\nu}}
\newcommand{\cJb}{\mathcal{J}_\mathrm{B}}
\newcommand{\cJ}{\mathcal{J}}

\newcommand{\vtrhd}{\mathcal{V}_\mathrm{tr}^\mathrm{HD}}
\newcommand{\vtrssep}{\mathcal{V}_\mathrm{tr}^\mathrm{1DSSEP}}
\newcommand{\vtrssept}{\mathcal{V}_\mathrm{tr}^\mathrm{2DSSEP}}
\newcommand{\phd}{p^\mathrm{HD}}
\newcommand{\qhd}{q^\mathrm{HD}}
\newcommand{\epsilonhd}{\epsilon^\mathrm{HD}}
\newcommand{\thd}{t^\mathrm{HD}}

\newcommand{\bV}{\mathbf{V}}
\newcommand{\bF}{\mathbf{F}}
\newcommand{\bxi}{\mathbf{\xi}}
\newcommand{\bP}{\mathbb{P}}

\begin{document}

\title{Driven tracers in narrow channels}

 \author{J. Cividini}
 \email{julien.cividini@weizmann.ac.il}
 \affiliation{Department of physics of complex systems,
 Weizmann Institute of Science\\
 Rehovot, Israel 76100}

 \author{D. Mukamel}
 \email{david.mukamel@weizmann.ac.il}
 \affiliation{Department of physics of complex systems,
 Weizmann Institute of Science\\
 Rehovot, Israel 76100}

\author{H.A. Posch}
\email{Harald.Posch@univie.ac.at}
\affiliation{Computational Physics Group, Faculty of Physics, 
 Universit\"at Wien,
 Boltzmanngasse 5, 1090 Vienna, Austria}

\date{\today}

\begin{abstract}
Steady state properties of a driven tracer moving in a narrow two dimensional (2D) channel of quiescent medium are studied.
The tracer drives the system out of equilibrium, perturbs the density and pressure fields, and gives the bath particles a nonzero average velocity, creating a current in the channel. Three models in which the confining effect of the channel is probed are analyzed and compared in this study: the first is the simple symmetric exclusion process (SSEP), for which  the stationary density profile and the pressure on the walls in the frame of the tracer are computed. We show that the tracer acts like a dipolar source in an average velocity field. The spatial structure of this 2D strip is then simplified  to a one dimensional SSEP, in which exchanges of position between the tracer and the bath particles are  allowed. Using a combination of mean field theory and exact solution in the limit where no exchange is allowed, gives good predictions of the velocity of the tracer and the density field. Finally, we show that results obtained for the 1D SSEP with exchanges also apply to a gas of overdamped hard disks in a narrow channel. The correspondence between the parameters of the SSEP and of the gas of hard disks is systematic and follows from simple intuitive arguments. Our analytical results are checked numerically.
\end{abstract}

\maketitle


\section{Introduction}

The influence of a driven particle (tracer) on the steady-state properties of the medium within which it is moving has been a subject of considerable experimental and theoretical interest in recent years. 
Driven tracers have been studied experimentally in a wide range of set\-ups  such as colloids dragged in DNA solution~\cite{gutsche2008}, spheres dragged in a polymer coil solution~\cite{kruger_r2009} or granular systems~\cite{candelier_d2010, pesic2012},  probe particles inside a colloidal crystal which locally melt the crystal~\cite{reichhardt_o2004,dullens_b2011}, or falling spheres in a fluid medium~\cite{chen1998, kahle_w_h2003, singh_s_g2012}. Questions of interest are, for example,  the steady state tracer velocity, the force-velocity relation, the local density distribution of the medium, the current induced by the tracer, and fluctuations of the tracer.

On the theoretical side, various approaches have been applied for studying driven tracers. They range from deterministic continuum hydrodynamic equations to 
models with stochastic dynamics such as field-theoretic path-integral approaches for the study of tracer diffusion \cite{demery_d2011}, Random Average Processes which provide analytical results for the density profile~\cite{cividini2016a, cividini2016b},  and a variety of discrete lattice gas models~\cite{burlatsky1992, burlatsky1996, landim_o_v1998, benichou1999, deconinck_o_m1997, benichou2001, benichou2000b, benichou2015, brummelhuis_h1989, illien2013, illien2015, benichou2013a,benichou2013b,benichou2013c}. 
The latter have proved rather useful for analyzing features such as density profiles, force-velocity relations, effect of geometrical constraints and fluctuations and correlations of the driven tracer. 
Extensive studies of a tracer subjected to a constant force have been carried out within the framework of the simple symmetric exclusion process (SSEP). In these models
the bath particles hop symmetrically on a lattice while the tracer is biased to preferentially hop in a particular direction. In addition, the bath particles may also undergo nonconserving adsorption-desorption processes. Infinite 1D~\cite{burlatsky1992, burlatsky1996, landim_o_v1998, benichou1999}, 2D~\cite{deconinck_o_m1997, benichou2001}, 3D spaces~\cite{benichou2000b} and even comb-like geometries~\cite{benichou2015} have been analyzed.  In a similar setup a tracer moving with a constant velocity has been studied in 2D using an Ising-like model \cite{demery_d2010a}. The case of an infinite one-dimensional line without absorption-desorption processes is a special case, where the particles stay ordered, so that the velocity of the tracer vanishes in the stationary state, asymptotically behaving like $t^{-1/2}$. In other cases, in particular in higher dimensions, the stationary velocity of the tracer is finite~\cite{deconinck_o_m1997, benichou2001,benichou2000b} and has been found, without much surprise, to be linear with the force for small driving force . 

Beyond the force-velocity relation, the SSEP framework allows one to probe the full position distribution of the tracer
~\cite{brummelhuis_h1989, illien2013, benichou2013a, illien2015}. In Ref.\,\cite{benichou2013a} it is shown that the position distribution converges to a Gaussian distribution. Its variance may however exhibit anomalous growth depending on the geometry~\cite{benichou2013a,benichou2013b,benichou2013c}.
In particular, for a quasi-1D narrow channel it has been shown that at large densities the position distribution of a tracer in a symmetric lattice gas converges to a Gaussian with variance $\simeq t^{3/2}$, a strongly superdiffusive behavior. In Ref.\,\cite{benichou2013a} this behavior has been linked to the covering properties of the random walk in this same geometry. The confinement indeed creates strong time correlations in the bath particles' density field.

In the present work we study the steady state properties of a tracer moving in a narrow channel. The confined environment is expected to have a strong effect on the properties of the gas, as has already been shown in equilibrium~\cite{forster_m_p2004,mukamel_p2009} and out of equilibrium in problems involving one~\cite{reguera_r2001, reguera2006, burada2007} or two particles~\cite{bowles_m_p2004}. Our study is carried out using three different models: The first  model is SSEP, where the simplicity of the dynamics allows one to calculate the steady state density and pressure profiles along the channel. The profiles show a peak ahead of the tracer with a dip behind it, and the velocity of the tracer is found to depend on $\Ly$. We also use a method for measuring the pressure in lattice gas models~\cite{dickman1987} for calculating the local pressure at the boundary to obtain the pressure profile in the framework of the moving tracer. The model is then further simplified by introducing a corresponding one-dimensional (1D) SSEP type
 model where, to reproduce the effect of the rows parallel to the one of the tracer, overtakes between bath particles and the tracer are allowed. This model is simpler to analyze and its steady state density profile is exactly calculable in some limit. The results obtained for the 1D model are readily compared with the ones of the 2D channel. In the third approach we consider a molecular dynamics model of overdamped hard disks (HD) in a narrow 2D channel. We show that the density profile and the velocity of the tracer in the hard disks model can be predicted to some extent using the results of the previous section for the 1D SSEP with overtakes. 

The paper is organized as follows. In Section~\ref{section:2dssep} we study the 2D SSEP model. We present analytical predictions from the discrete equations as well as from a simpler continuous equation and compare them to numerical results for the density field and the pressure. We then turn to the 1D SSEP with overtakes in Section~\ref{section:1dssep}. For this model approximate expressions for the current and the density profile are obtained. In Section~\ref{section:harddisks} the HD model is considered, and we show that a correspondence can be made between this more complicated model and the 1D SSEP. Section~\ref{section:sum} summarizes and concludes the paper.

\section{SSEP with a driven tracer in a two-dimensional narrow channel}
\label{section:2dssep}

\subsection{Model}

\begin{figure}[!ht]
	\begin{center}
		\includegraphics[width=0.45\textwidth]{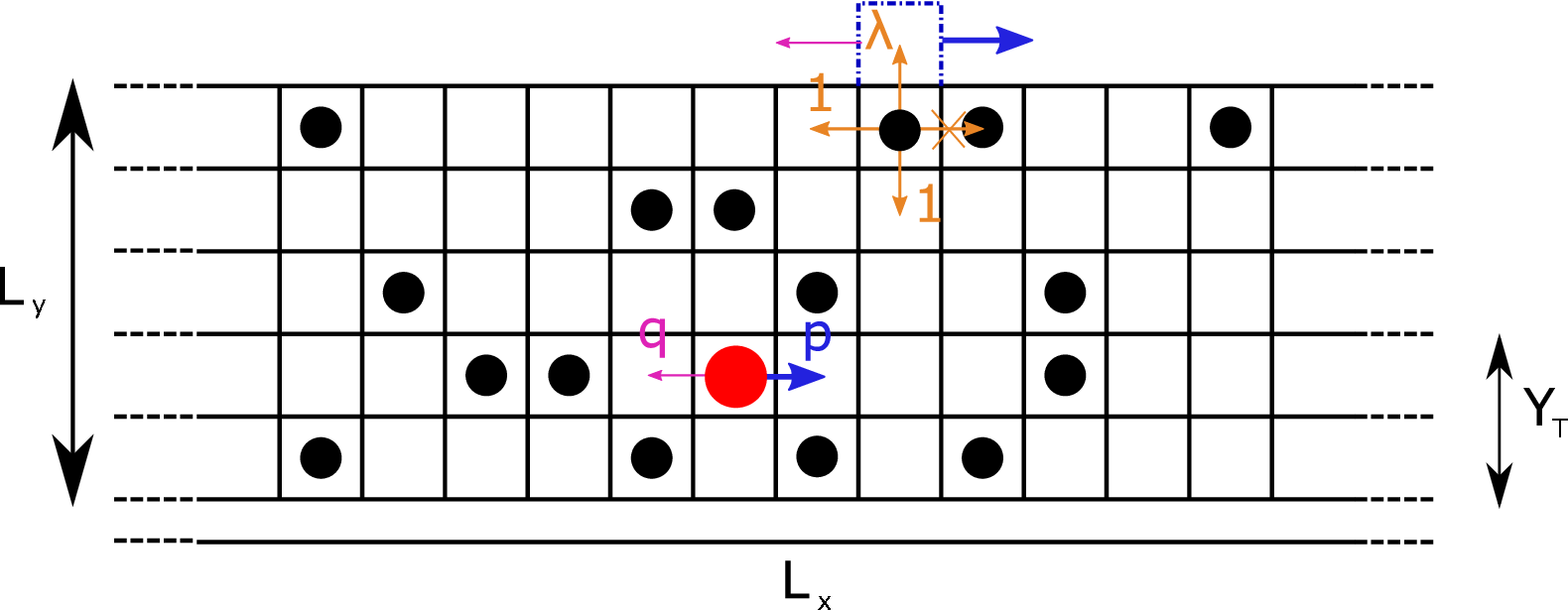} 
	\end{center}
	\caption{\small Scheme of the system studied for large $\Lx$, $\Ly=5$ and $\yT=2$.  The black disks are bath particles that may hop towards neighboring sites with rate $1$ if they are empty. The red disk is the tracer particle that hops with rate $p$ to the right and $q$ to the left. In order to measure the pressure, we introduce an extra site at the boundaries towards which bath particles are allowed to hop with rate $\lambda < 1$, (blue dashed-dotted site, also see subsection~\ref{subsubsection:defp}). The extra site always moves with the tracer, so that their distance is constant. In this picture the extra site is in a position to measure the pressure at a distance $(2,3)$ from the tracer.}  
	\label{fig:system}
\end{figure}

We start with studying the system drawn in Fig.\,\ref{fig:system}. We consider a two dimensional square lattice of length $\Lx$ in the $X$ direction and $\Ly$ in the $Y$ direction. We impose periodic boundary conditions in the $X$ direction and reflecting ones in the $Y$ direction. In the \textit{lattice frame} sites are denoted by $\bR = (X,Y)$ with $X=1,\ldots,\Lx$ and $Y=1,\ldots,\Ly$, and the basis vectors are denoted by $\bex$ and $\bey$. We consider the case of large $\Lx$ while keeping $\Ly$ small (of order $1$), which means that the particle will move in a narrow channel.

On this lattice we place one special particle, the \textit{tracer}, and $N$ identical \textit{bath} particles. The global density of the bath particles is $\rhoz \equiv \frac{N}{\Lx \Ly-1}$. The hard core exclusion constraint is enforced, \textit{i.e.} the maximum number of particles on each site is $1$. The model evolves by random sequential dynamics, so that time is continuous. Bath particles attempt to hop in each of the four directions with rate $1$. Their move is accepted if and only if their target site is empty. The bath particles are simply SSEP particles and would reach an equilibrium state with uniform distribution if there were no tracer.

In the most general case, the tracer would be allowed to move towards the four directions of space with different probabilities that depend on the force applied. 
In the present study we simplify the dynamics and allow the tracer to hop only in the $X$ direction keeping its $Y$ coordinate fixed.
We therefore allow the tracer to attempt to hop only in the $+X$ direction with rate $p$ and in the $-X$ direction with rate $q$. The position of the tracer in the lattice frame is denoted by $\brT = (\xT,\yT)$. By symmetry, we consider only the case $p>q$. In the following we analyze the steady state properties of the model.

\subsection{Analytical results}

\subsubsection{Equation for the density}

We begin by writing an equation for the density field in the frame of the tracer. A closed equation is obtained by factorizing the two-point correlations. The equation obtained here is a particular case of Eqs.\,(13)-(14) in Ref.\,\cite{benichou2001} on the full 2D plane. Here we however propose a different, somewhat shorter derivation and we apply it to the narrow channel geometry.

We denote positions in the frame of the tracer by $\br = (x,y)$, with 
\begin{eqnarray}
\label{eq:coordtracer}
x &=& X - \xT, \nonumber \\
y &=& Y-\yT.
\end{eqnarray}
 In this frame the tracer is fixed at position $\br = \bz$. Let us now define the occupation variables $\btau = \{\tau_\br\}$, that are $0$ for an empty site and $1$ for a site occupied by a bath particle, and write an equation for their evolution in time.

Examination shows that, for $\br \neq \bz, \pm \bex, \pm \bey$ we have 
\begin{equation}
\label{eq:taudiff}
\tau_\br(t+\dd t) - \tau_\br(t) =\Gamma_\br(t),
\end{equation}
with
\begin{equation}
\label{eq:noiseocc}
\Gamma_\br = 
\begin{cases}
 \tau_{\br+\benu}  &\text{w.~p.~}~ (1-\tau_\br) \dd t,~~\benu = \pm \bex, \pm \bey \\
 -\tau_{\br}  &\text{w.~p.~}~ (1-\tau_{\br+\benu}) \dd t,~~\benu = \pm \bex, \pm \bey\\
 \tau_{\br+\bex}-\tau_{\br}  &\text{w.~p.~}~p (1-\tau_\bex) \dd t\\
 \tau_{\br-\bex}-\tau_{\br}  &\text{w.~p.~}~q (1-\tau_{-\bex}) \dd t\\
\end{cases},
\end{equation}
where we have abbreviated 'with probability' by 'w.p.'. The first two lines are the usual SSEP terms and do not involve a motion of the tracer. The two last lines correspond to hops of the tracer. For example, the third line corresponds to the tracer hopping to the right between $t$ and $t+\dd t$ (probability $p (1-\tau_\bex) \dd t$ ), thus shifting site $\br+\bex$ into site $\br$.

The density is the ensemble average of the occupations, $\rho_\br(t) = \langle \tau_\br (t) \rangle$. The ensemble average of Eq.\,\eqref{eq:noiseocc} gives
\begin{widetext}
\begin{eqnarray}
\label{eq:occensavg}
\frac{\rho_\br(t+\dd t)-\rho_\br(t)}{\dd t} &=& \sum_{\benu = \pm \bex, \pm \bey } [\rho_{\br+\benu}(t)-\rho_\br(t)] + p [\rho_{\br+\bex}(t)-\rho_\br(t) - \langle \tau_{\br+\bex} \tau_\bex \rangle(t)+\langle \tau_{\br} \tau_\bex \rangle(t)] \nonumber \\
&&+ q [\rho_{\br-\bex}(t)-\rho_\br(t) - \langle \tau_{\br-\bex} \tau_{-\bex} \rangle(t)+\langle \tau_{\br} \tau_{-\bex} \rangle(t)].
\end{eqnarray}
\end{widetext}
Closed equations can be obtained for the density if one assumes that the pair correlations factorize, which is expected to be increasingly accurate at large $|\br|$. Repeating the same procedure for sites $\pm \bex, \pm \bey$ and factorizing the correlations, once again we get special equations for these sites. Defining 
\begin{eqnarray}
\label{eq:defApm}
\Ap &=& 1+p (1-\rho_{\bex}), \nonumber \\
\Am &=& 1+q (1-\rho_{-\bex}),
\end{eqnarray}
the equation for the density field at position $\br \neq \bz$ can be written as
\begin{eqnarray}
\label{eq:density}
\frac{\dd \rho_\br}{\dd t} &=& \Ap (\rho_{\br+\bex} -\rho_\br) + \Am (\rho_{\br-\bex} -\rho_\br) \nonumber \\ && + \rho_{\br+\bey} + \rho_{\br-\bey} - 2 \rho_\br \\
&& +\delta_{\br,\bex} (\Ap \rho_\bex - \Am \rho_\bz) + \delta_{\br,-\bex} (\Am \rho_{-\bex} -\Ap \rho_\bz) \nonumber \\ && + \delta_{\br,\bey} (\rho_\bey-\rho_\bz)  +\delta_{\br,-\bey} (\rho_{-\bey}-\rho_\bz). \nonumber 
\end{eqnarray}
The boundary conditions are periodic in the $x$ direction, $x \equiv x + \Lx$, 
 and reflecting in the $y$ direction,
\begin{eqnarray}
\label{eq:densitybcy}
\rho_{(x,-\yT+1)} &=& \rho_{(x,-\yT)}, \nonumber \\
\rho_{(x,\Ly-\yT)} &=& \rho_{(x,\Ly-\yT+1)}.
\end{eqnarray}
For $\Ly=2 \yT-1$ the tracer is in the middle of the channel and the two equations ~\eqref{eq:densitybcy} become equivalent by symmetry. In this case, reflecting boundary conditions in the $y$ direction become equivalent to periodic boundary conditions.
Equations~\eqref{eq:density} involve an auxiliary field $\rho_\bz$. 
Its value does not matter, as it cancels out in the equations for the physical variables at $\br \neq \bz$, but it has been kept with the purpose of regularizing Eqs.\,\eqref{eq:density} at $\br = \bz$, see below. It can be checked from Eq.\,\eqref{eq:density} that mass is conserved, $\frac{\dd }{\dd t} \sum_{\br \neq \bz} \rho_\br = 0$. As already stated, the bulk equation \eqref{eq:density} is a special case of Eqs.\,(13)-(14) in Ref.\,\cite{benichou2001} when absorption, desorption and hopping of the tracer in the $y$ direction all vanish, and after rescaling of the time. In Ref.\,\cite{benichou2001} $\rho_\bz$ was chosen equal to $0$. The boundary conditions are however different from Ref.\,\cite{benichou2001}, where the whole plane is studied.

Since $\rho_\bz$ is arbitrary, one may add it as an auxiliary variable which satisfies $\frac{\dd \rho_\bz}{\dd t} = 0$. Eqs.\,\eqref{eq:density} may then be generalized to include the equation for $\rho_\bz$, yielding
\begin{eqnarray}
\label{eq:densityf}
\frac{\dd \rho_\br}{\dd t} &=& \Ap (\rho_{\br+\bex} -\rho_\br) + \Am (\rho_{\br-\bex} -\rho_\br) \nonumber \\ && + \rho_{\br+\bey} + \rho_{\br-\bey} - 2 \rho_\br \nonumber \\
&& +(\delta_{\br,\bex}-\delta_{\br,\bz}) (\Ap \rho_\bex - \Am \rho_\bz) \nonumber \\ && + (\delta_{\br,-\bex}-\delta_{\br,\bz}) (\Am \rho_{-\bex} -\Ap \rho_\bz) \\
&&+ (\delta_{\br,\bey}-\delta_{\br,\bz}) (\rho_\bey-\rho_\bz) \nonumber \\ && +(\delta_{\br,-\bey}-\delta_{\br,\bz}) (\rho_{-\bey}-\rho_\bz). \nonumber
\end{eqnarray}
We note that mass on all sites of the channel including $\bz$ is conserved, $\frac{\dd }{\dd t} \sum_{\br} \rho_\br = 0$. 

For simplicity, let us consider a tracer in the middle of the channel and solve for the stationary state. In order to demonstrate qualitatively the behavior of the system under study, we simplify~\eqref{eq:densityf} by considering an analogous continuous version of it. Choosing $\rho_\bz=0$, we obtain
\begin{eqnarray}
\label{eq:rhoc}
(\Ap - \Am) \partial_x \rho &+& \frac{\Ap+\Am}{2} \partial_x^2 \rho + \partial_y^2 \rho \\ &=& (\Ap \rho_\bex - \Am \rho_{-\bex}) \partial_x \delta(\br), \nonumber
\end{eqnarray} 
which is valid in a narrow channel $-\infty < x < \infty$, $-\frac{\Ly}{2} < y < \frac{\Ly}{2}$ with the reflecting boundary conditions in the $y$ direction, $\partial_y \rho |_{y=\pm \frac{\Ly}{2}} = 0$.
The discrete density field $\rho_\br$ has been replaced by a coarse-grained version $\rho(\br)$ that is allowed to take any real value. 
 Besides the diffusion terms $\frac{\Ap+\Am}{2} \partial_x^2 \rho + \partial_y^2 \rho$, in~\eqref{eq:rhoc} we kept the advection term $(\Ap - \Am) \partial_x \rho$ resulting from the fact that we are in a moving frame,  and the dipolar source term $(\Ap \rho_\bex - \Am \rho_{-\bex}) \partial_x \delta(\br)$.

The effect of a \textit{driven bond} (rather than a tracer) on an infinite square lattice was studied in Refs.\,\cite{sadhu_m_m2011,sadhu_m_m2014b}. Comparing the evolution equation~\eqref{eq:densityf} with Eq.\,(9) of Ref.\,\cite{sadhu_m_m2011}, one notices that the evolution of the density is the same as in a system of SSEP particles moving on a lattice with four driven bonds, which drive the particles between $\bz$ to the four neighboring sites. In Ref.\,\cite{sadhu_m_m2011} it has been shown that a driven bond produces a density perturbation similar to the potential produced by a dipole at large distances.

 The dipolar source term appearing on the RHS of Eq.\,\eqref{eq:rhoc} results from the combination of the $+\bex$ and $-\bex$ source terms in the second line of~\eqref{eq:densityf}. In the case of a tracer driven in the $x$ direction in the middle of the channel, the $+\bey$ and $-\bey$ source terms have the same magnitude and opposite directions. Their dipolar contributions therefore compensate on large scales and the sum of the $+\bey$ and $-\bey$ terms contributes to a higher, quadrupolar order. Similarly to a driven bond and a dipole, a driven tracer creates an accumulation of bath particles at its front and a depletion at its back. 

Contrary to the driven bond problem, the nonzero velocity of the tracer gives rise to an advection term in equation~\eqref{eq:rhoc}. The presence of this term screens the long-range character of the dipolar field, as shown in Ref.\,\cite{sadhu_m_m2011}. 

\begin{figure}[!ht]
	\begin{center}
		\includegraphics[width=0.45\textwidth]{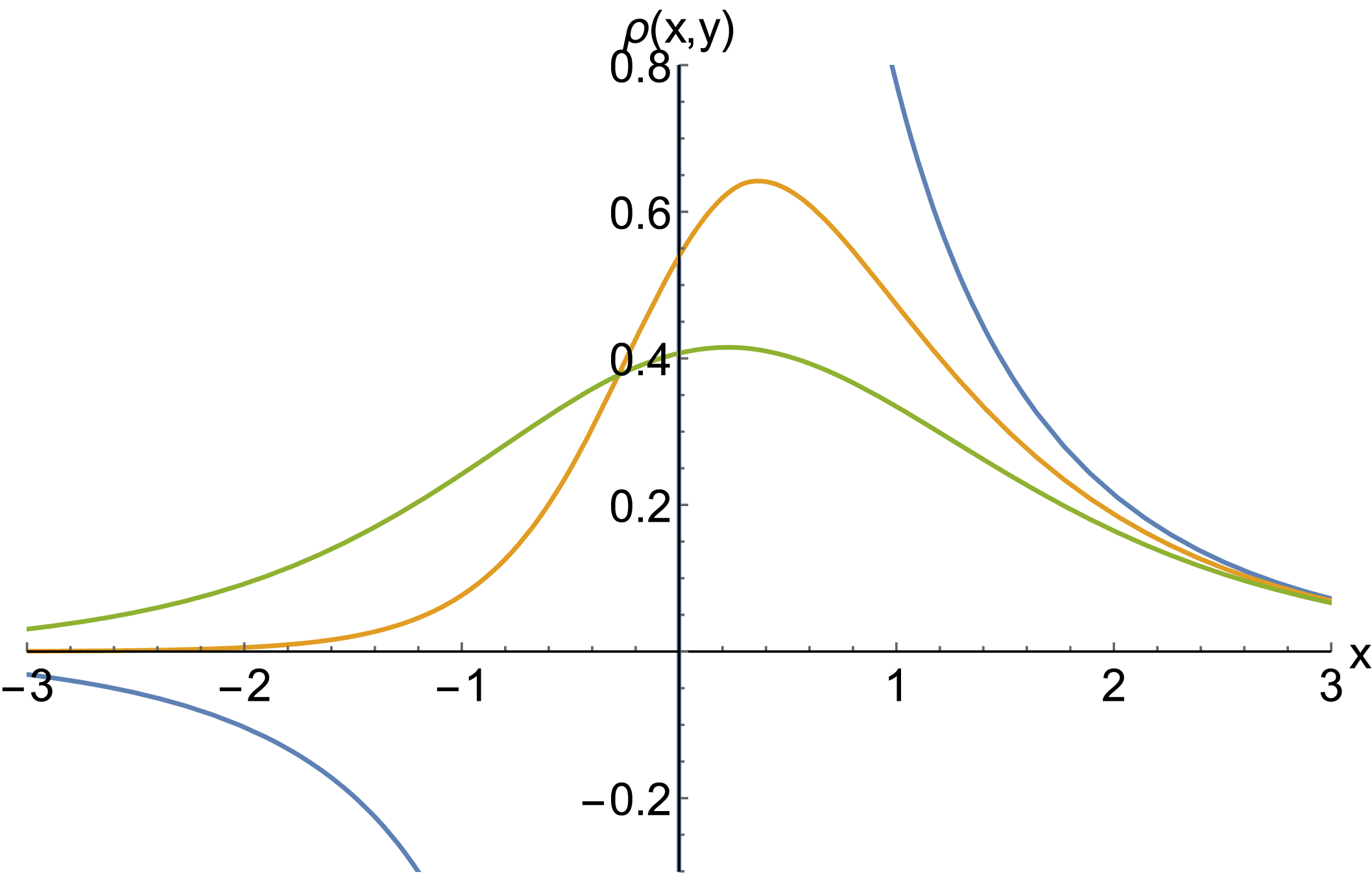} 
	\end{center}
	\caption{\small Solution~\eqref{eq:rhosolc} of the continuous equation~\eqref{eq:rhoc} for $\Ly=4$, $\rhoz = 0$ and $\Ap - \Am = \frac{\Ap+\Am}{2} = \Ap \rho_\bex - \Am \rho_\bey = 1$. Each curve is for a constant value of $y=0,1$ or $2$, for the blue, orange and green curve, respectively. 
		}  
	\label{fig:densitycont}
\end{figure}

In the narrow-channel case the solution is obtained by taking a Fourier transform in the $x$ direction and expanding in cosines in the $y$ direction,
\begin{widetext}
\begin{equation}
\label{eq:tfc}
\rho(\br) = \rhoz + \left[ \int_{k=-\infty}^\infty a_0(k) \ee^{2 \pi \ci k x} \dd k \right]  + 2 \sum_{m=1}^\infty \left[ \int_{k=-\infty}^\infty a_m(k) \ee^{2 \pi \ci k x} \dd k \right] \cos \left( \frac{2 m \pi}{\Ly} y\right),
\end{equation}
which satisfies \eqref{eq:rhoc} for
\begin{equation}
\label{eq:solam}
a_m(k) = - \frac{\Ap \rho_\bex - \Am \rho_{-\bex}}{\Ly} \frac{2 \pi \ci k}{\frac{\Ap+\Am}{2} 4 \pi^2 k^2 +\frac{4 m^2 \pi^2}{\Ly^2} - (\Ap - \Am) 2 \pi \ci k}.
\end{equation}
After performing the $k$ integrals we get
\begin{equation}
\label{eq:rhosolc}
\rho(\br) = \rhoz + \frac{2(\Ap \rho_\bex - \Am \rho_{-\bex})}{\Ly (\Ap+\Am)} \left[ \Theta(x) \ee^{-\frac{x}{\xi}} + \sum_{m=1}^\infty (\sgn x +\frac{1}{r_m}) \ee^{- \frac{1+ r_m \sgn x}{2} \frac{x}{\xi}} \cos \left( \frac{2 m \pi}{\Ly} y\right) \right] ,
\end{equation}
\end{widetext}
where we defined $\xi = \frac{\Ap+\Am}{2 (\Ap-\Am)}$ and $r_m = \sqrt{1+  \frac{2(\Ap+\Am)}{(\Ap-\Am)^2}\frac{4 \pi^2}{\Ly^2} m^2}$, and $\Theta(x)$ is the Heaviside step function. As is shown in  Fig.\,\ref{fig:densitycont}, the solution~\eqref{eq:rhosolc} is continuous everywhere except at $\br = \bz$, where it diverges.

When $x>0$, for each value of $m$ the corresponding factor in Eq.\,\eqref{eq:rhosolc} decays exponentially in $x$ with a characteristic length $\frac{2 \xi}{1+r_m}$. At large distances, the first term due to $m=0$ dominates and the decay length, $\xi$, stays finite in the $\Ly \to \infty$ limit. This decay can be attributed to the effective streaming in the frame of the tracer. When $x < 0$, the $m=0$ term vanishes. The other terms decay exponentially with $x$ with a characteristic distance $\frac{2 \xi}{r_m-1}$, which diverges like $\frac{ (\Ap-\Am) \Ly^2}{4 m^2 \pi^2 }$ when $\Ly$ is large and $m$ finite. 
In the limit $\Ly \rightarrow \infty$, the decay is again exponential everywhere except for $y=0$. In this latter case, the cosines in the sum are all $1$, the decay length at finite $\Ly$ is given by $\frac{2}{r_{1}-1}$, and in the $\Ly \rightarrow \infty$ limit the decay length diverges to give the $|x|^{-3/2}$ behavior observed on the plane. The solution decays exponentially in all directions except in the full plane $\Ly \to \infty $ at the back of the tracer/dipole (for $x < 0$ and $y=0$), as expected from the results presented in Refs.\,\cite{benichou2001, sadhu_m_m2011} and the discussion following Eq.
\eqref{eq:rhoc}.

For later comparison, we also study the density of particles projected on the $x$ axis. Integrating over $y$, only the first two terms of Eq.\,\eqref{eq:rhosolc} remain, and we obtain
\begin{equation}
\label{eq:rhoxc}
\int_{y=-\frac{\Ly}{2}}^\frac{\Ly}{2} \rho(x,y) \dd y = \Ly \rhoz + \frac{2(\Ap \rho_\bex - \Am \rho_{-\bex})}{\Ap+\Am} \Theta(x) \ee^{-\frac{x}{\xi}}.
\end{equation}
There is an accumulation of particles in front of the tracer in a region thin in the $y$ direction, while the projected density at the back of the tracer is unperturbed.

The scale of this density perturbation is given by the strength of the dipole. For weak driving the dipolar moment goes like $2(\rho_\bex \Ap - \rho_{-\bex} \Am) \simeq 2 (p-q) \rhoz (1-\rhoz)$. Intuitively, the perturbation should indeed vanish at an empty or fully occupied channel, and is quite reasonably proportional to the drive. In the narrow channel case, it is expected that the density in the vicinity in front of the tracer increases when the channel becomes narrower, as the excess particles in front have less room to make way for the tracer. The continuous solution~\eqref{eq:rhosolc} is physically and mathematically easy to understand, but now needs to be compared to the solution of the discrete problem.

\subsubsection{Solution of the discrete system for a finite channel width}

It is of interest to go beyond the continuum approximation~\eqref{eq:rhoc} and to study the steady state of the discrete system explicitly. To this end we solve Eq.\,\eqref{eq:density} in the stationary state. As the value of $\rho_\bz$ does not matter, we may as well choose $\rho_\bz = 0$. We proceed as if the amplitudes of the source terms, which actually depend on $\rho_{\pm \bex}$ and $\rho_{\pm \bey}$, were known. In that case Eq.\,\eqref{eq:density} becomes linear and can therefore be solved separately for each source term. More precisely, we can write
\begin{eqnarray}
\label{eq:densitysolfor}
\rho_\br &=& \rhoz + \Ap \rho_\bex (G_{\br|\bex}-G_{\br|\bz})  + \Am \rho_{-\bex} (G_{\br|-\bex}-G_{\br|\bz}) \nonumber \\ && + \rho_\bey (G_{\br|\bey}-G_{\br|\bz}) + \rho_{-\bey} (G_{\br|-\bey}-G_{\br|\bz}),
\end{eqnarray}
where the Green's function $G_{\br|\br'}$ is the solution of
\begin{eqnarray}
\label{eq:G}
\Ap (G_{\br+\bex|\br'} -G_{\br|\br'}) + \Am (G_{\br-\bex|\br'} -G_{\br|\br'}) && \\ + G_{\br+\bey|\br'} + G_{\br-\bey|\br'} - 2 G_{\br|\br'} &=& -\delta_{\br,\br'}. \nonumber
\end{eqnarray}
It implicitly depends on $\yT$ through the boundary conditions,
\begin{eqnarray}
\label{eq:Gbc}
G_{(x,-\yT+1)|\br'} &=& G_{(x,-\yT)|\br'}, \nonumber \\
G_{(x,\Ly-\yT)|\br'} &=& G_{(x,\Ly-\yT+1)|\br'}.
\end{eqnarray}

To solve equation Eq.\,\eqref{eq:G}, we take Fourier transforms  in both directions, 
\begin{eqnarray}
\label{eq:defft}
\hG_{(q_x,y)|\br'} &\equiv& \sum_{x=1}^\Lx G_{\br|\br'} \ee^{-\frac{2 \pi \ci q_x x}{\Lx}}, \\
\tG_{\bq|\br'} &\equiv& \sum_{x=1}^\Lx \sum_{y=-\yT+1}^{\Ly-\yT} G_{\br|\br'} \ee^{-\frac{2 \pi \ci q_x x}{\Lx}-\frac{2 \pi \ci q_y y}{\Ly}} \nonumber \\ &=& \sum_{y=-\yT+1}^{\Ly-\yT} \hG_{(q_x,y)|\br'} \ee^{-\frac{2 \pi \ci q_y y}{\Ly}}, \nonumber
\end{eqnarray}
where $\bq = (q_x, q_y)$.
The inverse transformations are given by
\begin{eqnarray}
\label{eq:defftinv}
 \hG_{(q_x,y)|\br'} &=& \Ly^{-1} \sum_{q_y = 0}^{\Ly-1} \tG_{\bq|\br'} \ee^{\frac{2 \pi \ci q_y y}{\Ly}}, \\
 G_{\br|\br'} &=& \Lx^{-1} \sum_{q_x = 0}^{\Lx-1} \hG_{(q_x,y)|\br'} \ee^{\frac{2 \pi \ci q_x x}{\Lx}} \nonumber \\ &=& (\Lx \Ly)^{-1} \sum_{q_x = 0}^{\Lx-1} \sum_{q_y = 0}^{\Ly-1} \tG_{\bq|\br'} \ee^{\frac{2 \pi \ci q_x x}{\Lx}+\frac{2 \pi \ci q_y y}{\Ly}}. \nonumber
\end{eqnarray}
 Defining 
 \begin{eqnarray}
 \label{eq:Lambda}
 \Lambda_\bq &\equiv& \Ap (\ee^{\frac{2 \pi \ci q_x}{\Lx}}-1)+\Am (\ee^{-\frac{2 \pi \ci q_x}{\Lx}}-1) \nonumber \\ &&+ \ee^{\frac{2 \pi \ci q_y}{\Ly}} + \ee^{-\frac{2 \pi \ci q_y}{\Ly}} - 2,
 \end{eqnarray}
 we get from Eq.\,\eqref{eq:G} 
\begin{eqnarray}
\label{eq:Gft}
&& \Lambda_\bq \tG_{\bq|\br'} + \ee^{-\frac{2 \pi \ci q_x}{\Lx} x' -\frac{2 \pi \ci q_y}{\Ly} y'}  \\ &=& (\hG_{(q_x,\Ly-\yT)|\br'}- \hG_{(q_x,-\yT+1)|\br'}) \ee^{\frac{2 \pi \ci q_y}{\Ly} \yT} (\ee^{-\frac{2 \pi \ci q_y}{\Ly}}-1).\nonumber
\end{eqnarray}
The first term on the right-hand side (RHS) comes from the delta source,  the second term from the boundaries.  In particular, if the tracer is in the middle of the channel, \textit{i.e.} $\Ly=2 \yT-1$, these terms vanish by symmetry. Equation \eqref{eq:Gft} can be solved for $\tG_{\bq|\br'}$ except for $\bq=\bz$. The $\bq=\bz$ term however only leads to a constant in the final expression of $G_{\br|\br'}$, which has already been accounted for in Eq.\,\eqref{eq:densitysolfor}. We therefore use the convention $\tG_{\bz|\br'} = 0$ in the following. After solving for $\tG_{\bq|\br'}$ we transform back in the $y$ direction only and get
\begin{eqnarray}
\label{eq:Gftx}
\hG_{(q_x,y)|\br'} &=& -\hcA_{(q_x,y)|\br'} \\ && + (\hG_{(q_x,\Ly-\yT)|\br'}- \hG_{(q_x,-\yT+1)|\br'}) \hcB_y, \nonumber 
\end{eqnarray} 
where we define
\begin{eqnarray}
\label{eq:defAB}
\hcA_{(q_x,y)|\br'} &\equiv& \Ly^{-1} \sum_{q_y=0}^{\Ly-1} \frac{\ee^{-\frac{2 \pi \ci q_x}{\Lx} x' -\frac{2 \pi \ci q_y}{\Ly} (y'-y)}}{\Lambda_\bq}, \nonumber \\
\hcB_y &\equiv& \Ly^{-1} \sum_{q_y=0}^{\Ly-1} \frac{\ee^{\frac{2 \pi \ci q_y}{\Ly} (\yT+y)} (\ee^{-\frac{2 \pi \ci q_y}{\Ly}}-1)}{\Lambda_\bq}.
\end{eqnarray}
The value of $\hG_{(q_x,\Ly-\yT)|\br'}- \hG_{(q_x,-\yT+1)|\br'}
 $ can be determined self-consistently, and we finally obtain
\begin{eqnarray}
\label{eq:Gfinal}
G_{\br|\br'} &&= - \cA_{\br|\br'} \\
+ \Lx^{-1} \sum_{q_x=0}^{\Lx-1} &&\frac{\hcA_{(q_x,-\yT+1)|\br'}-\hcA_{(q_x,\Ly-\yT)|\br'}}{1+\hcB_{-\yT+1}-\hcB_{\Ly-\yT}} \hcB_y \ee^{\frac{2 \pi \ci q_x x}{\Lx}},\nonumber
\end{eqnarray}
where 
\begin{equation}
\label{eq:defA}
\cA_{\br|\br'} \equiv \Lx^{-1} \sum_{q_x=0}^{\Lx-1} \hcA_{(q_x,y)|\br'} \ee^{\frac{2 \pi \ci q_x x}{\Lx}}
\end{equation}
 is the bulk term and the most important in magnitude. From equations \eqref{eq:densitysolfor} and \eqref{eq:Gfinal} it can be shown that the density perturbation decays algebraically at the back of the tracer only for $\Lx$ and $\Ly$ both infinite, and exponentially everywhere else. This is consistent with the continuous case~\eqref{eq:rhosolc}, where a true $|x|^{-3/2}$ decay is obtained in the $l \to \infty$ limit only.

Equations \eqref{eq:densitysolfor} and \eqref{eq:Gfinal} do not constitute a full solution of Eq.\,\eqref{eq:densityf}, as the densities $\rho_\bex, \rho_{-\bex}, \rho_\bey, \rho_{-\bey}$ should now be determined self-consistently. This is, however, a very hard task, and in the following we choose to measure these quantities numerically and to take them as inputs. The numerical results will also be compared with the theoretical expression \eqref{eq:densitysolfor}-\eqref{eq:Gfinal}, where the densities close to the tracer are approximated by their zeroth-order values,
\begin{equation}
\label{eq:approxrho}
\rho_\bex \simeq \rho_{-\bex} \simeq \rho_\bey \simeq \rho_{-\bey} \simeq \rhoz,
\end{equation}
which also gives
\begin{equation}
\label{eq:approxA}
\Ap \simeq 1+p (1-\rhoz), \qquad \Am \simeq 1+q (1-\rhoz).
\end{equation}

 Similar to~\eqref{eq:rhoxc}, one can compute the number of particles at a given $x$ coordinate starting from the discrete solution~\eqref{eq:densitysolfor}-\eqref{eq:Gfinal}. We get
 \begin{equation}
 \label{eq:decaysumdiscr}
 \rho_x - \Ly \rhoz \simeq \Theta(x) \left( \rho_\bex \frac{\Ap}{\Am} - \rho_{-\bex} \frac{\Am}{\Ap} \right) \left( \frac{\Am}{\Ap} \right)^x, 
 \end{equation}
when the channel is long in the $x$ direction. The density, integrated over $y$, again shows an exponentially decaying perturbation for $x>0$, and remains unchanged for $x<0$.

\subsection{Comparison with Monte-Carlo simulations}

\subsubsection{Density profile}

\begin{figure}[!ht]
	\centering
	\begin{subfigure}[b]{0.41\textwidth}
		\centering
		\includegraphics[width=\textwidth]{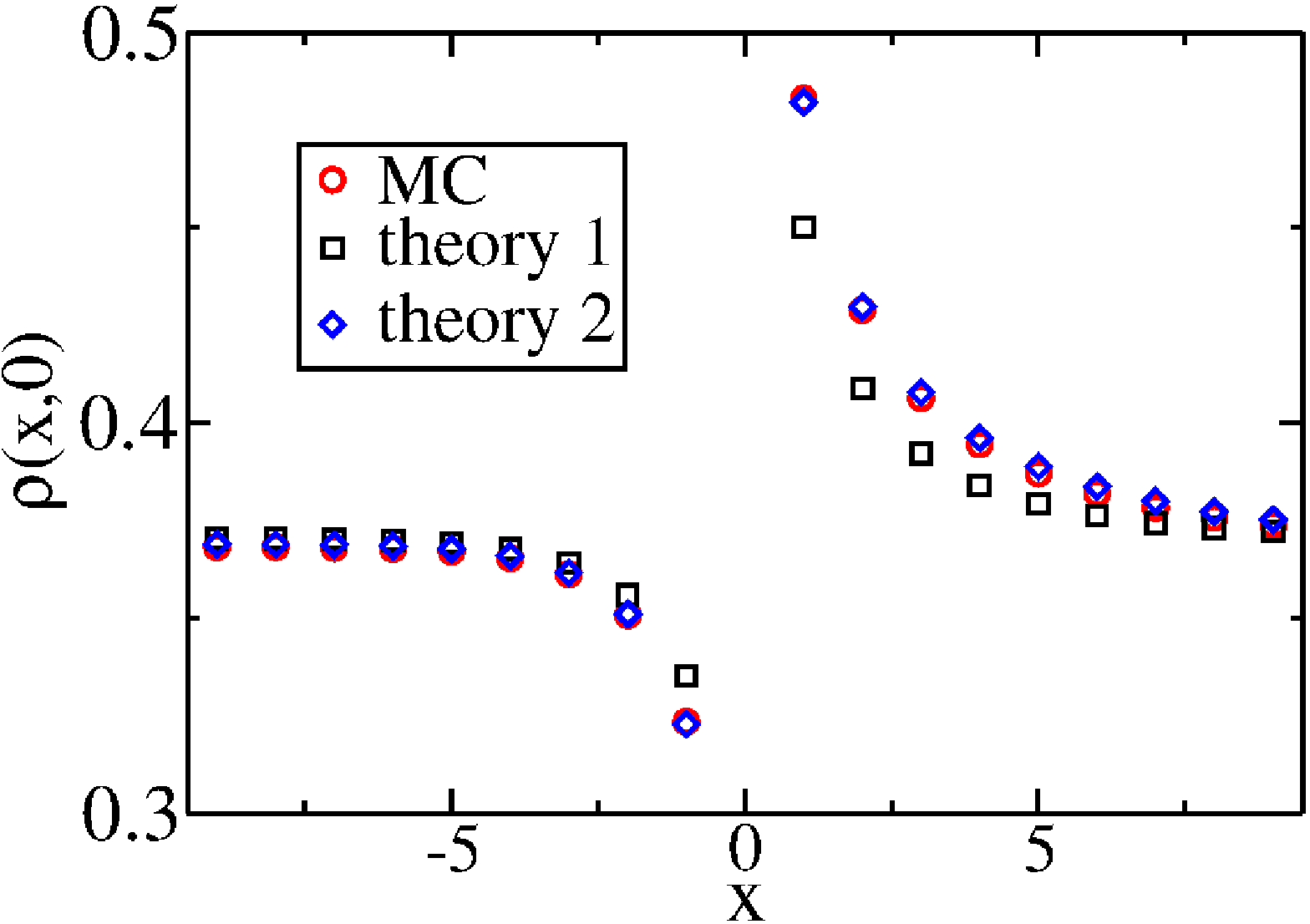}
		\caption{$y=0$}
		\label{fig:densityprofiley0}
	\end{subfigure}
	\hfill
	\begin{subfigure}[b]{0.41\textwidth}
		\centering
		\includegraphics[width=\textwidth]{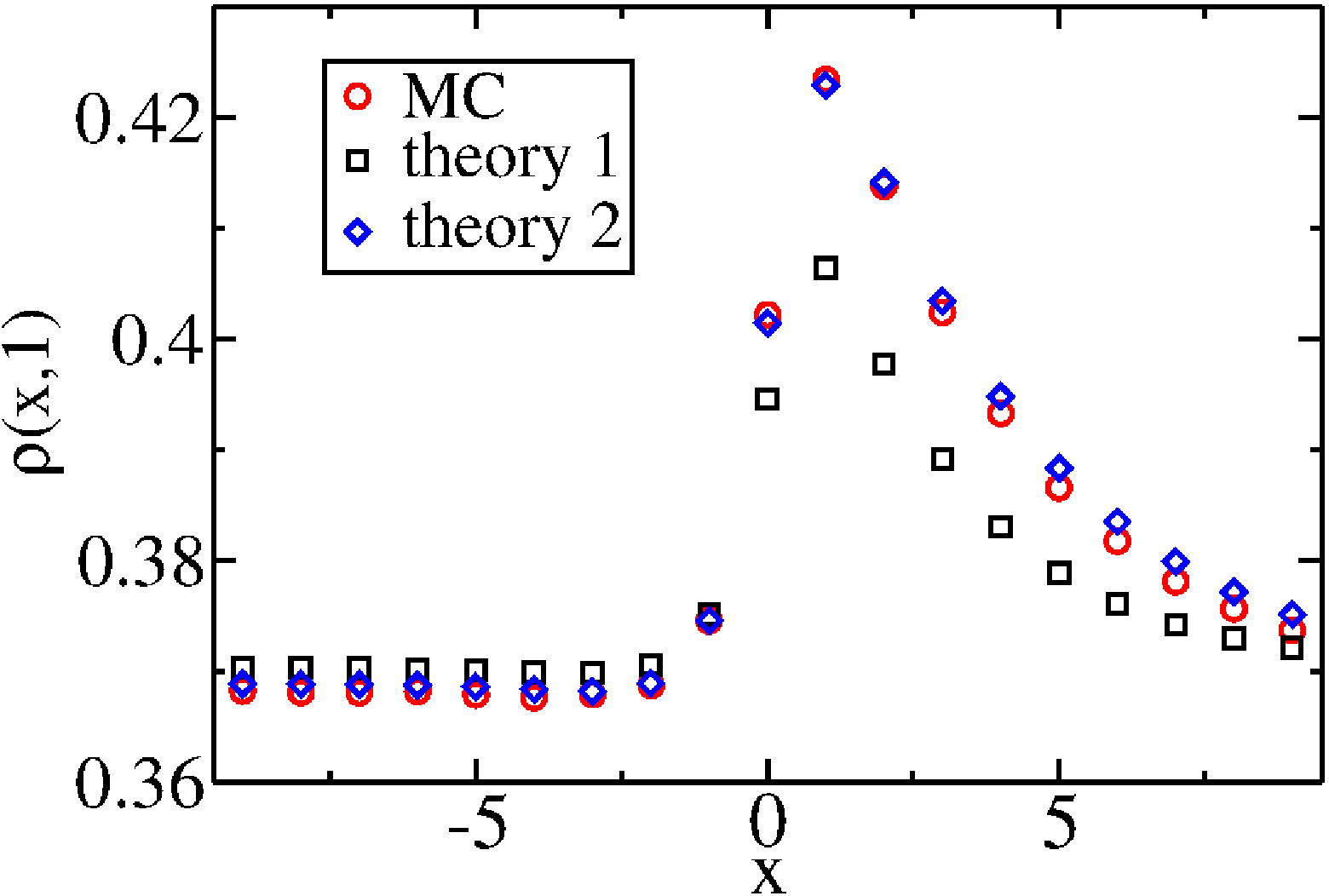}
		\caption{$y=1$}
		\label{fig:densityprofiley1}
	\end{subfigure}
	\hfill
	\begin{subfigure}[b]{0.41\textwidth}
		\centering
		\includegraphics[width=\textwidth]{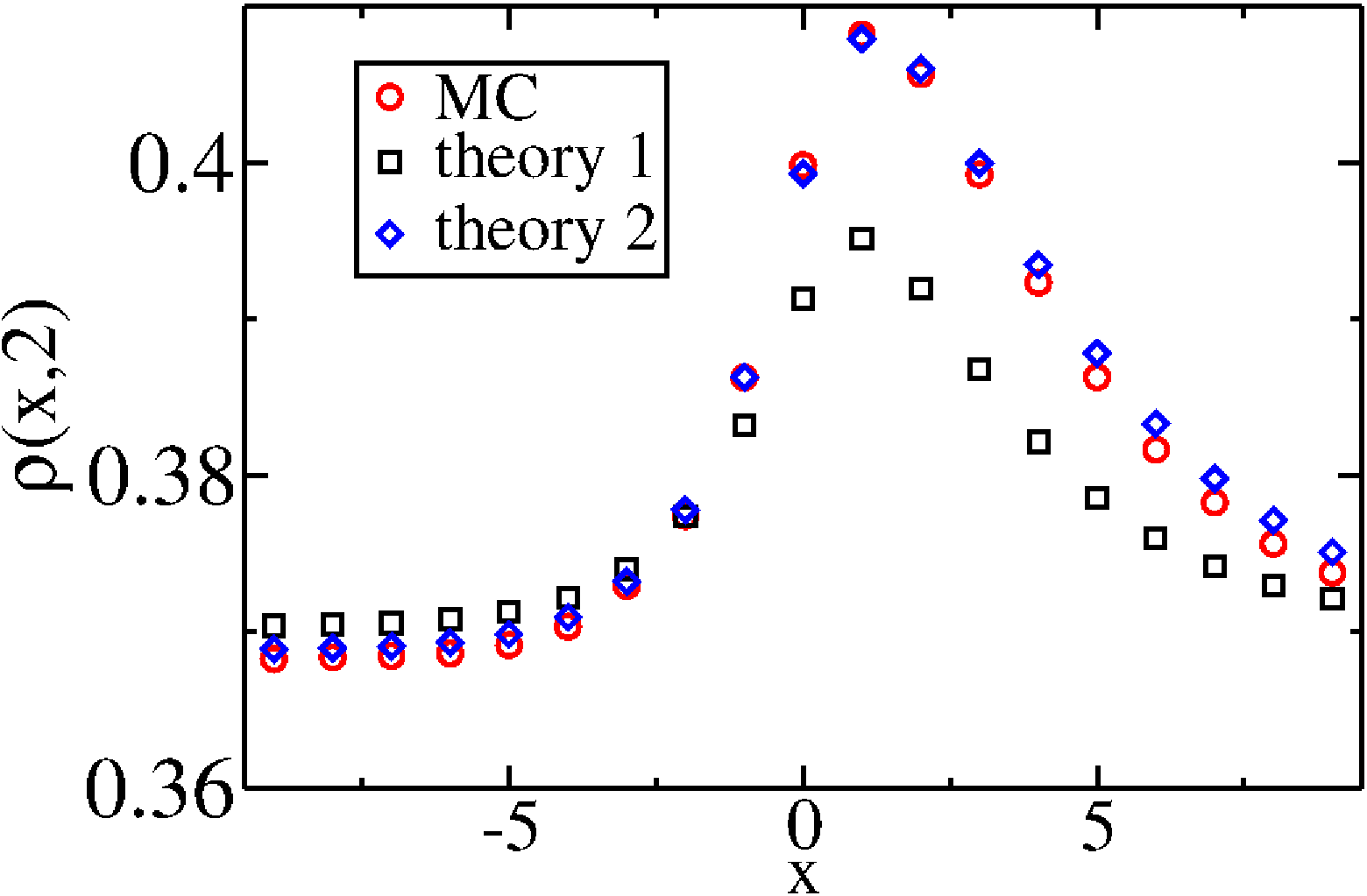}
		\caption{$y=2$}
		\label{fig:densityprofiley2}
	\end{subfigure}
	\caption{\small Density profile along the $x$ direction on each row for a thin channel with $\Ly=5$ for fixed $y=0$ (a), $y=1$ (b) and $y=2$ (c). On each graph the results of Monte-Carlo (MC) simulations are plotted with red circles, the theoretical expression \eqref{eq:densitysolfor}-\eqref{eq:Gfinal} using the approximations \eqref{eq:approxrho}-\eqref{eq:approxA} with black squares (theory 1), and the theoretical expression using the numerically measured values of $\rho_\benu$, $\benu = \pm \bex, \pm \bey$ \eqref{eq:densitysolfor}-\eqref{eq:Gfinal}  with blue diamonds (theory 2). The other parameters are $\Lx=81$, $N=150$, $p=1.5$, $q=0.5$, and the tracer is in the middle of the channel, $\yT = 3$. For these parameters the measured values are $\rho_\bex=0.48$, $\rho_{-\bex}=0.32$, $\rho_\bey=\rho_{-\bey}=0.40$ and $\Ap = 1.77$, $\Am=1.33$, whereas \eqref{eq:approxrho}-\eqref{eq:approxA} give $\rho_\bex=\rho_{-\bex}=\rho_\bey=\rho_{-\bey}=\frac{150}{81 \times 5 -1} \simeq 0.37$, $\Ap=1.94$ and $\Am=1.31$.}
	\label{fig:densityprofile}
\end{figure}

We now compare the analytical and the numerical results for a driven tracer. We start by noting that in the symmetric equilibrium case, $p=q$, detailed balance is restored, and the equilibrium distribution is uniform over all the allowed configurations. Therefore, the pressure and density are constant even in the frame of the tracer. This can be checked numerically and is in accordance with the theoretical expression~\eqref{eq:densitysolfor}-\eqref{eq:Gfinal}, where the four sources add up to a constant. We now focus on the asymmetric case $p \neq q$.

We start by placing the tracer particle in the center of the channel, $\Ly = 2 \yT - 1$. Measurements of the density profile are averaged over a sufficiently long time, typically $10^6$ time units. They are shown in Fig.\,\ref{fig:densityprofile} on the row of the tracer in panel (a), and on the two first neighboring rows $y=1$ in panel (b) and $y=2$ in panel (c). There we show two theoretical expressions, one where the $\rho_\benu$, $\benu = \pm \bex, \pm \bey$ are determined  numerically, and another one where Eqs.\eqref{eq:approxrho}-\eqref{eq:approxA} are used.  When the measured values of $\rho_\benu$, $\benu = \pm \bex, \pm \bey$ are used, the agreement is quantitatively very good, and it stays qualitatively correct if  the zeroth-order approximation \eqref{eq:approxrho}-\eqref{eq:approxA} is applied.
Fig.\,\ref{fig:densityprofiley0} shows that there is a discontinuity of the density profile along the row $y=0$ at the position of the tracer, $x=0$. In the neighboring rows the excess density for $x > 0$ gets transported with the effective flow created by the motion of the tracer and progressively fills the region $x < 0$ as $|x|$ increases. The only zone that persistently stays depleted is the one at the back of the tracer, which is not filled ballistically but diffusively by the particles of adjacent rows.

\begin{figure}[!ht]
	\begin{center}
		\includegraphics[width=0.45\textwidth]{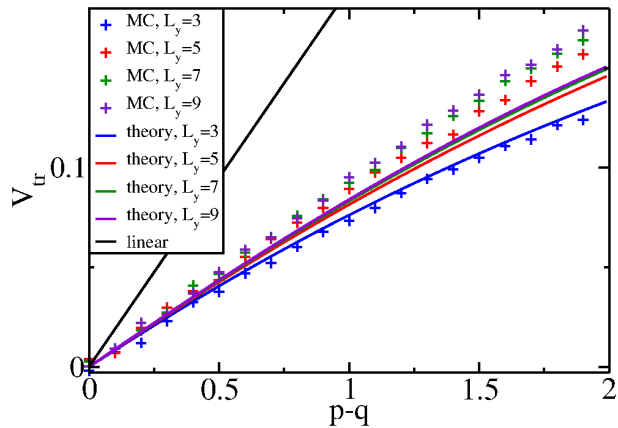} 
	\end{center}
	\caption{\small Tracer velocity $\vtrssept$ as a function of the bias $p-q$ for different system sizes. Monte-Carlo measurements (pluses) are compared to theory~\eqref{fig:vtr2ssep}-\eqref{eq:densitysolfor}-\eqref{eq:Gfinal} (solid lines) and to the linear approximation $\vtrssept = (p-q)(1-\rhoz)$ (black solid line). The length of the channel $\Lx=41$ and the global density is held approximately constant, $\rhoz \simeq 0.81$.}  
	\label{fig:vtr2ssep}
\end{figure}

In terms of the density, the velocity of the tracer is given by
\begin{equation}
\label{eq:vtr2ssep}
\vtrssept = p (1-\rho_{\bex}) - q(1-\rho_{-\bex}).
\end{equation}
It increases less than linearly with the bias $p-q$, in accordance with the fact that the tracer has to struggle against a higher density gradient when the bias is larger. The velocity shows a small decrease as the channel becomes narrower, since the front of the tracer becomes more crowded, see Fig.\,\ref{fig:vtr2ssep}.

The profiles show that the presence of a boundary results in an increased density of particles ahead of the tracer. The theoretical expression of Eqs. \eqref{eq:densitysolfor}-\eqref{eq:Gfinal} reproduces this effect rather well. 
The fact that the tracer is not  centered does not have a significant effect.

\subsubsection{Pressure profile}
\label{subsubsection:defp}

In this section we describe a numerical method for evaluating the local pressure on the walls of the channel. To this end, we apply a procedure proposed in Ref.\,\cite{dickman1987} in order to measure numerically the pressure of lattice gases at equilibrium in the homogeneous case. To describe the method, we first consider the homogeneous case with no tracer present, so that we are simply studying an SSEP in a channel of size $\Lx \times \Ly$. The stationary state of this system is an equilibrium state described by a partition function $Z(N,\Lx,\Ly)$ in the lattice frame. 

To measure the pressure on the $Y=\Ly$ boundary, one should carry out the following steps:
\begin{itemize}
	\item Choose a rate $\lambda \in [0;1]$ and define the system where hops from $Y=\Ly-1$ to $Y=\Ly$ occur with a rate $\lambda$, while all other rates remain unchanged equal to $1$.
	\item In this modified system, measure $\rho_W(\lambda)$, the average equilibrium occupation of the sites in the  row $\Ly$.
	\item Do this for all $\lambda \in [0;1]$.
\end{itemize}
To factors negligible in the thermodynamic limit, the rescaled pressure is given by~\cite{dickman1987}
\begin{eqnarray}
\label{eq:defpressuredickman}
\pr &\equiv& \Lx^{-1} [ \log Z(N,\Lx,\Ly) - \log Z(N,\Lx,\Ly-1) ] \nonumber \\
&=& \int_{\lambda=0}^{1} \frac{\rho_W(\lambda)}{\lambda} \dd \lambda.
\end{eqnarray}
Note that we have canceled out some factors that appear in the usual definition of the pressure. 
In order to make a correspondence between the lattice gas model and a system at temperature $T$ the rescaled pressure $P$ should be compared to $\frac{P^{\mathrm{nsc}}}{\kb T}$, where $P^{\mathrm{nsc}}$ is the true pressure in the system.

In our case we need to adapt the method,  since the pressure is not homogeneous at the wall, the stationary state is not an equilibrium state, and the measurements are performed in a moving frame, the frame of the tracer.
Suppose the tracer is on row $\yT$. The pressure at position $(x,\Ly-\yT)$ in the tracer frame is obtained by adding a site $(x,\Ly-\yT+1)$ towards which the hops occur with rate $\lambda$, see Fig.\,\ref{fig:system}. The stationary density on this site is determined and the integral in Eq.\,\eqref{eq:defpressuredickman} is computed. We expect that the pressure reaches a nontrivial steady state only in the frame of the tracer, therefore the extra site has to move with the tracer and stay at position $(x,\Ly-\yT+1)$ in the tracer frame. In the lattice frame, if the tracer hops from $\brT$ to $\brT \pm \bex$, then the extra site also moves from $(\xT+x,\Ly+1)$ to $(\xT+x+1,\Ly+1)$. If there is a particle on the extra site, we choose that the particle moves with the site.

\begin{figure}[!ht]
	\begin{center}
		\includegraphics[width=0.45\textwidth]{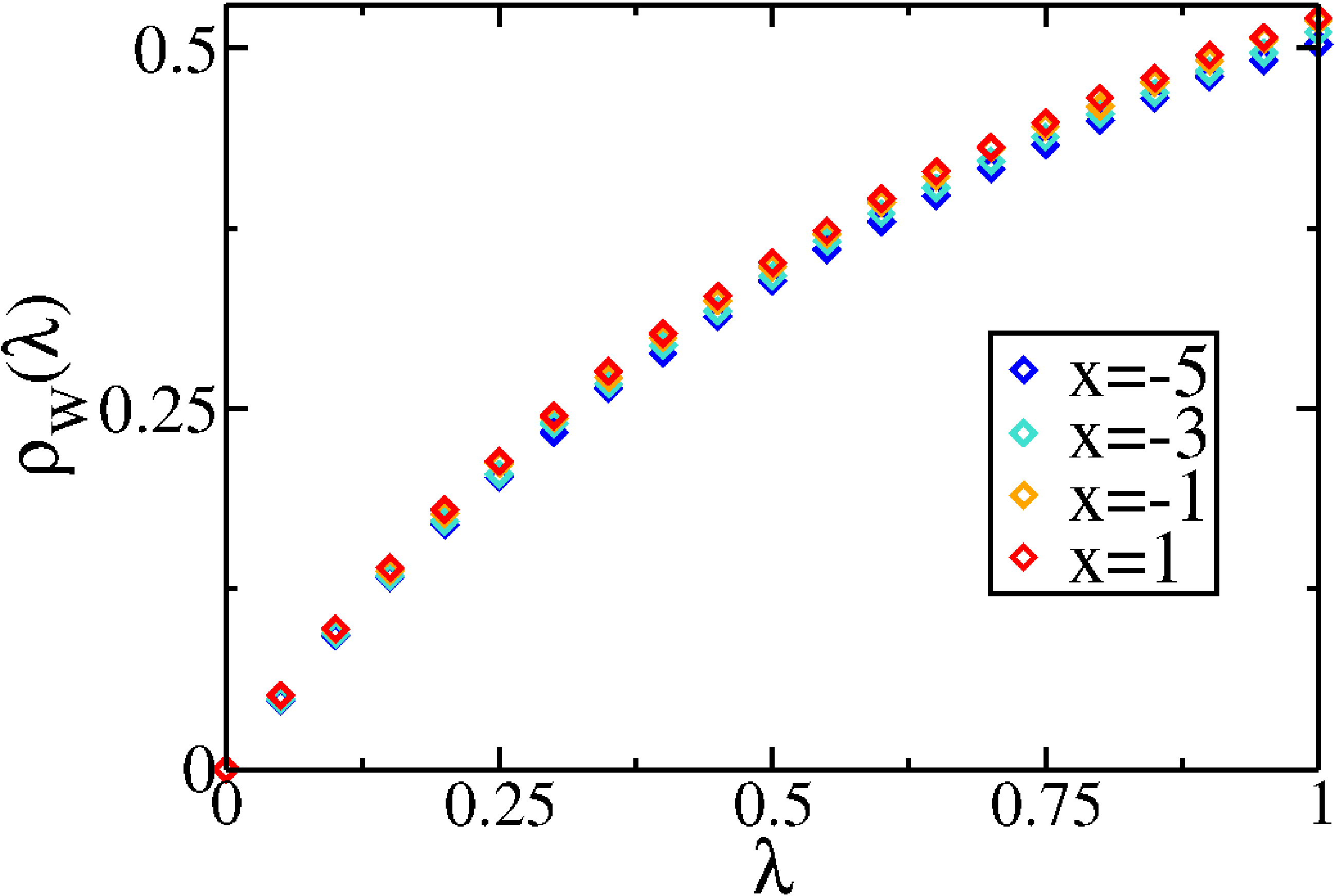} 
	\end{center}
	\caption{\small Measured $\rho_W(\lambda)$ for a system of size $\Lx=81$, $\Ly=9$,  with the tracer in the channel center, with $\yT = 5$. The pressure is obtained at a distance $(x,4)$ from the tracer, where the curves are shown for $x=-5$, $-3$, $-1$ and $1$. The hopping parameters of the tracer are $p=1.9$ and $q=0.1$, and there are $N=365$ particles in the system.}  
	\label{fig:rhow}
\end{figure}

\begin{figure}[!ht]
	\begin{center}
		\includegraphics[width=0.45\textwidth]{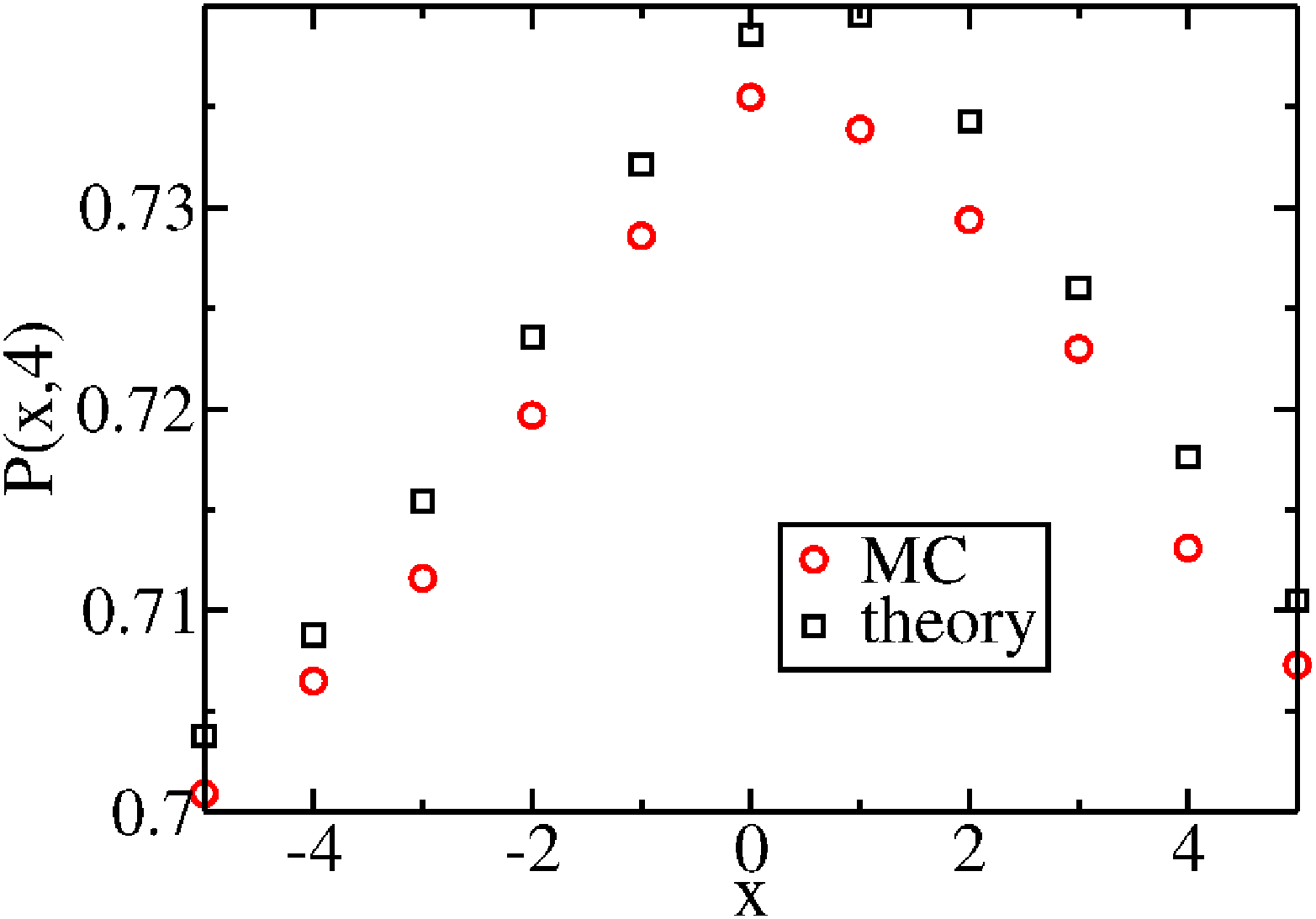} 
	\end{center}
	\caption{\small Monte-Carlo (blue) and theoretical (orange) values of the pressure for the same system as in Fig.\,\ref{fig:rhow}.  The value of the background pressure is $-\log(1-\rhoz) \simeq -\log\left(1-\frac{365}{9 \times 81-1}\right) \simeq 0.696$}  
	\label{fig:pressure}
\end{figure}

We have measured the pressure using this method and the density $\rho_W(\lambda)$ is given in Fig.\,\ref{fig:rhow}. The occupation of the extra site $\rho_W(\lambda)$ was averaged over a sufficiently long time for all values of $\lambda$ from $0$ to $1$ in multiples of $0.05$. The curve $\rho_W(\lambda)$ was then fitted by a seventh-order polynomial without constant coefficient, and the integral~\eqref{eq:defpressuredickman} was computed directly from the coefficients of the polynomial. The pressure curve is given in Fig.\,\ref{fig:pressure}.

The assumption of local equilibrium yields an alternative way for estimating the pressure. For SSEP the equilibrium equation of state is known, and the pressure simply becomes  $\pr=-\log(1-\rhoz)$, where $P$ is the (homogeneous) pressure and $\rhoz$ the global average density. In our case we assume local equilibrium, such that the pressure becomes 
\begin{equation}
\label{eq:pressuressep}
\pr_\br=-\log(1-\rho_\br),
\end{equation}
where $\pr_\br$ and $\rho_\br$ are the respective local pressure and density.
The pressure is then given by Eq.\,\eqref{eq:pressuressep} combined with the theoretical expression for the density in Eqs.\,\eqref{eq:densitysolfor}-\eqref{eq:Gfinal}. It compares well with the results from the other method, see Fig.\,\ref{fig:pressure}. Again the agreement is good. There is a systematic error of order $(\Lx \Ly)^{-1}$, which may be explained by two effects. First, when we move the extra site we may move a particle with it, which slightly changes the density profile. Second, we gradually add a site to the system, so that there is also a systematic error that comes from the determination of the exact value of the density.

Fig.\,\ref{fig:pressure} shows that the presence of the tracer creates a significant pressure perturbation for a typical density $\rhoz = \frac{365}{9 \times 81-1} \simeq 0.5$. In particular, it shows that on the sides the pressure is larger than in the equilibrium case. This comes form the fact that there is an accumulation of bath particles in front of the tracer that have to go around it to enable the tracer to move forward.

\section{One-dimensional SSEP with exchanges}
\label{section:1dssep}

In this section we further simplify the 2D model analyzed in the previous section and model the narrow channel by a 1D discrete exclusion process with a tracer in an SSEP background. The 2D nature of the narrow channel is taken into account by allowing the tracer to overtake the bath particles with some rate. The simplicity of this approach, for which exact steady-state density profiles can be computed in some limits,  allows for a quantitative comparison with the steady state obtained by molecular dynamics simulations of overdamped hard disks moving in a narrow channel, a model which will be analyzed in Section \ref{section:harddisks}.

\subsection{Model}

\begin{figure}[!ht]
	\begin{center}
		\includegraphics[width=0.45\textwidth]{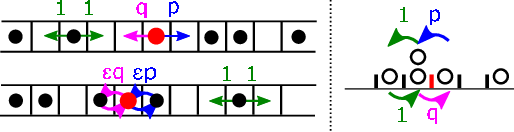} 
	\end{center}
	\caption{\small Left panel:  Two possible configurations of the system and some allowed transitions. The black disks are the bath SSEP particles, and the red disk is the tracer. Bath particles hop symmetrically towards right and left with rate $1$ on each side if their target site is empty. The tracer hops to the right with rate $p$ and to the left with rate $q$ if its target site is empty, as is shown by the top scheme. A tracer and a neighboring particle exchange their positions with rate $\epsilon p$ or $\epsilon q$, if the tracer moves to the right or to the left, respectively. This may occur in the configuration at the bottom of the left panel.\\
		Right panel:  Configuration of the Zero-Range Process (ZRP) equivalent to the top left SSEP for the case $\epsilon=0$. 
		The tracer is mapped to a special link (red tick), where the transfer rates are $q$ when the ZRP particle hops to the right, and $p$ when the particle hops to the left as indicated by the purple and blue arrows. 
		}  
	\label{fig:schemesystem}
\end{figure}

We consider a periodic 1D lattice of $\Lx$ sites occupied by $N$ SSEP particles with hard-core interactions hopping symmetrically to the right or left with rate $1$. To this system we add a driven tracer, that hops to the right with rate $p$ and to the left with rate $q$. The tracer is also allowed to exchange position with neighboring SSEP particles, with rate $\epsilon p$ to the right and $\epsilon q$ to the left (see Fig.\,\ref{fig:schemesystem}). When $\epsilon$ is not too small we expect the 1D SSEP introduced here to mimic the behavior of the 2D system from Section~\ref{section:2dssep} on a qualitative level.

We would like to compute the density profile in the frame of the tracer. We therefore use again the coordinate system where the tracer is on site $0$. The other particles may now occupy sites $1$ to $\Lx-1$, and occupations are denoted by $\tau_x =0,1$ for $x=1,\ldots,\Lx-1$. The average density now reads $\rhoz = \frac{1}{\Lx-1} \sum_{x=1}^{\Lx-1} \rho_x = \frac{N}{\Lx-1}$. Other quantities of interest are the bath particle current in the lattice frame $\cJb$, the velocity of the tracer $\vtrssep$, and the total (bath and tracer) particle current $\cJ$. The velocity of the tracer can be defined in terms of the density in the frame of the tracer,
\begin{eqnarray}
\label{eq:defvtr}
\vtrssep &=& p (1-\rho_1) + \epsilon p \rho_1 \nonumber \\ &&- q (1-\rho_{\Lx-1}) - \epsilon q \rho_{\Lx-1}.
\end{eqnarray}
In the steady state both currents $\cJb$ and $\cJ$ are space independent. The current of bath particles, $\cJb$, can also be computed  at any position $X$ of the lattice frame  in terms of the density profile in the tracer frame,
\begin{eqnarray}
\label{eq:defjb}
\cJb &=& \langle (1-\delta_{X,X_T}-\delta_{X+1,X_T})(\tau_X - \tau_{X+1})\rangle \nonumber \\ &&+\epsilon \langle p \tau_X \delta_{X+1,X_T}-q \tau_{X+1} \delta_{X,X_T}\rangle \nonumber \\
&=& \frac{(1-\epsilon p) \rho_1 -(1-\epsilon q) \rho_{\Lx-1}}{\Lx}.
\end{eqnarray}
The currents $\cJb$ and $\cJ$ and the tracer velocity $\vtrssep$ are linked through
\begin{eqnarray}
\label{eq:jjbvtr}
\cJ &=& \cJb + \frac{\vtrssep}{\Lx} \\ &=& \frac{p+\rho_1 (1- p) - q - \rho_{\Lx-1} (1- q)}{\Lx}, \nonumber
\end{eqnarray} 
which simply expresses the fact that the total current is the sum of the currents of the bath particles and of the tracer. 

We restrict the analysis to the case $p > q$ and $\epsilon < 1$ due to symmetries. The case $\epsilon > 1$ can be obtained by exchanging particles and holes, $\tau_x \rightarrow 1-\tau_x$, and replacing $p$, $q$ and $\epsilon$ by $\epsilon p$, $\epsilon q$ and $\epsilon^{-1}$. Systems with $p < q$ are obtained by reflection symmetry with respect to the direction $X$.

We start with some particular values of $\epsilon$ that give exactly known steady states. After that we approximately compute the density profile for general epsilon and large systems.

\subsection{Limiting cases}

\subsubsection{$\epsilon=0$}

For $\epsilon=0$ no exchanges of positions are allowed between the SSEP particles and the tracer. In this case, the stationary state of the system can be found exactly by mapping it to a Zero-Range Process (ZRP)~\cite{evans_h2005}. 
The ZRP is a very general process in which particles occupy sites of a graph and are allowed to hop from a site to one of its neighbors with a rate that depends only on the occupation of the starting site. A particle therefore interacts only with particles occupying the same site, hence the name 'zero range'. In our case it suffices to consider the ZRP on a 1D ring with $\lambda$ sites and $\nu$ particles. The number of particles on a given site $l$ is denoted by $n_l = 0,1,\ldots,\infty$, and a configuration of the system is given by $\bn = (n_1,\ldots,n_\lambda)$. 
 
The correspondence is as follows. Each vacancy of the SSEP is mapped to a particle of the ZRP, each particle of the SSEP is mapped to a link of the ZRP, while a site of the ZRP corresponds to an interval between two particles of the SSEP (see Fig.\,\ref{fig:schemesystem}). For the present model this gives $\lambda = N$ and $\nu = \Lx-N$. The occupation $n_l$ of site $l$ of the ZRP is equal to the number of vacancies between the particles corresponding to links $(l-1,l)$ and $(l,l+1)$. In our case all the links of the ZRP represent symmetric SSEP particles, except for one special link that we choose to be $(1,\lambda)$, that represents the tracer. Examination shows that the transition rates should be
\begin{eqnarray}
\label{eq:rateszrplneqlambda}
\bn \to& (\ldots,n_l-1,n_{l+1}+1,\ldots) ~~&\mathrm{rate}~~(1-\delta_{n_l,0}), \\
\bn \to& (\ldots,n_l+1,n_{l+1}-1,\ldots) ~~&\mathrm{rate}~~(1-\delta_{n_{l+1},0}), \nonumber
\end{eqnarray}
for any $l \neq \lambda$ and
\begin{eqnarray}
\label{eq:rateszrpleqlambda}
\bn \to& (n_1+1,\ldots,n_\lambda-1) ~~&\mathrm{rate}~~q (1-\delta_{n_\lambda,0}),  \\
\bn \to& (n_1-1,\ldots,n_\lambda+1) ~~&\mathrm{rate}~~p (1-\delta_{n_{1},0}). \nonumber
\end{eqnarray}
An example of this mapping in a particular configuration is shown in Fig.\,\ref{fig:schemesystem}. Note that carrying out this mapping for $\epsilon > 0$ would result in a similar process, where the special link would be able to hop under certain conditions, making it harder to treat.

The benefit of this mapping is that the stationary state of the ZRP is known exactly~\cite{evans_h2005} and has only to be translated to the SSEP variables. It takes a factorized form except for a global constraint,
\begin{equation}
\label{eq:pstate0}
\bP(\bn) = \frac{\prod_{l=1}^\lambda z_l^{n_l}}{\cZln} \delta_{\sum_l n_l, \nu},
\end{equation}
where 
\begin{equation}
\label{eq:zzrp}
\cZln = \sum_{\bn} \prod_{l=1}^\lambda z_l^{n_l} \delta_{\sum_l n_l, \nu}
\end{equation}
is the normalization constant, and the \textit{fugacities} $z_l$ are solutions of
\begin{eqnarray}
\label{eq:rwzl}
q z_\lambda + z_2 &=& (1+p) z_1 \\
z_{l-1} + z_{l+1} &=& 2 z_l \qquad l=2,\ldots,\lambda-1 \nonumber \\
z_{\lambda-1} + p z_1 &=& (1+q) z_\lambda, \nonumber
\end{eqnarray} 
that is,
\begin{equation}
\label{eq:solrwzl}
z_l = \frac{l}{\lambda} + \frac{q \lambda -p+1}{\lambda (p-q)}.
\end{equation}

Back to real space, let $x$ be a site of the SSEP in the tracer frame. Site $x$ is occupied if and only if there exists site $m$ of the ZRP such that $\sum_{l=1}^m n_l + m = x$. We therefore have
\begin{eqnarray}
\label{eq:rhoe0}
\cZln \rho_x &=& \sum_{m=1}^\infty \cZln \langle \delta_{\sum_{l=1}^m n_l + m,x} \rangle \nonumber \\
&=& \sum_{m=1}^\infty \sum_{\bn} \delta_{\sum_l n_l, \nu} \delta_{\sum_{l=1}^m n_l + m,x} \prod_{l=1}^\lambda z_l^{n_l} \nonumber \\
&=& \sum_{m=1}^\infty \frac{1}{2 \pi \ci} \oint_u \frac{\dd u}{u^{\nu+1}} \frac{1}{2 \pi \ci} \oint_v \frac{\dd v}{v^{x-m+1}} \nonumber \\
&& \times \prod_{l=1}^m \frac{1}{1-u v z_l} \prod_{l=m+1}^\lambda \frac{1}{1-u z_l} \nonumber \\
&\simeq& \lambda \int_{\mu=0}^\infty \frac{1}{2 \pi \ci} \oint_u \frac{\dd u}{u} \frac{1}{2 \pi \ci} \oint_v \frac{\dd v}{v} \ee^{-\lambda \phi(u,v,\mu)},\nonumber
\end{eqnarray}
where on the third line the Kronecker deltas have been represented by complex integrals. The fourth line is valid in the large $\lambda$ limit with the scaling $\nu = r \lambda$, $x = \xi \lambda$ and $m = \mu \lambda$, where
\begin{eqnarray}
\label{eq:defphi}
\phi(u,v,\mu) &=& r \log u + (\xi-\mu) \log v \nonumber \\ &&+ \int_{y=0}^\mu \log (1-u v (y + c)) \dd y  \\ &&+ \int_{y=\mu}^1 \log (1-u (y + c)) \dd y, \nonumber
\end{eqnarray}
and $c = \frac{q}{p-q}$.  $r$ is linked to the density via $r = \rhoz^{-1}-1$. In the large $\lambda$ limit one can find a saddle point in the $(u,v,\mu)$ complex planes, where  $\vst = 1$, $\must = \frac{1-\ust c}{\ust} (1-\ee^{-\ust \xi})$, and $\ust$ is defined implicitly by 
\begin{equation}
\label{eq:ustimpl}
(1+r) \ust + \log \left( \frac{1-\ust (1+c)}{1- \ust c} \right) = 0.
\end{equation}
It can be shown that Eq.\,\eqref{eq:ustimpl} has a unique real root that lies between $\frac{1+2 c -\sqrt{(1+2c)^2 -4 \frac{r}{1+r} c (1+c)}}{2 c (1+c)}$ and $\frac{1}{1+c}$. For small bias $p-q$ we have
\begin{eqnarray}
\label{eq:ustexp}
\ust &=& 2 (1-\rhoz) \frac{p-q}{p+q} - \frac{2 (1-\rhoz)^3}{3 \rhoz} \left( \frac{p-q}{p+q}\right)^3 \nonumber \\ &&+ O\left(\left( \frac{p-q}{p+q}\right)^5\right).
\end{eqnarray}

We can evaluate~\eqref{eq:rhoe0} using a saddle point approximation, and perform a similar calculation to obtain $\cZln$. Since $\vst = 1$, almost all the terms that appear cancel out with the $\cZln$ factor on the LHS, and we are left with
\begin{equation}
\label{eq:rhoe0f}
\rho_x = \rho_{\lambda \xi} =(1-\ust c) \ee^{-\ust \xi} = (1-\ust c) \ee^{-\frac{\ust}{N} x}.
\end{equation} 
In this case the profile is purely exponential, and the decay length is of order $N$.

The velocity of the tracer is obtained by computing the average densities in Eq. \eqref{eq:defvtr},
\begin{equation}
\label{eq:vtreps0}
\vtrssep = \langle q (1-\delta_{n_N,0}) - p (1-\delta_{n_1,0}) \rangle = \frac{\ust}{N}.
\end{equation}
Since no overtake is possible for $\epsilon=0$, all the particles have to move with the same velocity, \textit{i.e.} the velocity of the tracer. Since the tracer only contributes a fraction $1/N$ to the total current, the total and the bath-particle currents are the same to leading order in $1/N$, and both become
\begin{equation}
\label{eq:jbjeps0}
\cJb = \cJ = \rhoz \vtrssep = \frac{\ust}{\Lx}.
\end{equation}

In a similar manner, it can be shown by explicit calculation that the correlations between occupations factorize in the $\Lx \rightarrow \infty$ limit.

\subsubsection{$\epsilon=1$}

Another particularly simple case is $\epsilon = 1$, since for $\epsilon=1$ the tracer does not distinguish between particles and holes. 
We denote the probability to have a certain occupation by $\bP(\btau)$, where $\btau = (\tau_1,\ldots,\tau_{\Lx-1})$. We also define $\btau^{x,x+1}$, where $\tau_x$ and $\tau_{x+1}$ have been exchanged, $\tau^\rightarrow = (\tau_{\Lx-1},\tau_1,\ldots,\tau_{\Lx-2})$ and $\tau^\leftarrow = (\tau_2,\ldots,\tau_{\Lx-1},\tau_1)$. The master equation reads
\begin{eqnarray}
\label{eq:mastere1}
\frac{\dd \bP(\btau)}{\dd t} &=& \sum_{x=1}^{\Lx-2} [(1-\tau_x) \tau_{x+1} \bP(\btau^{x,x+1}) - \tau_x (1-\tau_{x+1}) \bP(\btau)] \nonumber \\
&&\hspace{-5mm}+ \sum_{x=2}^{\Lx-1} [(1-\tau_x) \tau_{x-1} \bP(\btau^{x,x-1}) - \tau_x (1-\tau_{x-1}) \bP(\btau)] \nonumber \\ &&\hspace{-5mm}+ p [\bP(\btau^\rightarrow)-\bP(\btau)] + q [\bP(\btau^\leftarrow)-\bP(\btau)].
\end{eqnarray}
The terms of the first and second lines correspond to hops of the bath particles, and the terms of the third line come from the motion of the tracer. It is clear that a constant $\bP(\btau) = \binom{\Lx-1}{N}^{-1}$ solves~\eqref{eq:mastere1} in the stationary state. The density profile is therefore flat, and all the correlations are the same as in an SSEP of length $\Lx-1$ with $N$ particles, \textit{e.g.}
\begin{eqnarray}
\label{eq:corre1}
\langle \tau_x \tau_{x'} \rangle &=& \frac{N}{\Lx-1} \frac{N-2}{\Lx-2} \\
&=& \rho_x \rho_{x'} - \frac{N(\Lx-N-1)}{(\Lx-1)^2 (\Lx-2)}, \nonumber 
\end{eqnarray} 
and the connected part vanishes as $\Lx^{-1}$ for large systems. 

Using the flat uncorrelated density profile the currents are quite easy to obtain. Since the tracer moves to the right with rate $p$ and to the left with rate $q$ regardless of the occupations, the velocity of the tracer is simply
\begin{equation}
\label{eq:vtreps1}
\vtrssep = p-q.
\end{equation}
The only transitions that contribute to the bath particle current are those where the tracer exchanges position with a bath particle. The current is given by the intuitive result
\begin{equation}
\label{eq:jbeps1}
\cJb = -\frac{(p-q) \rhoz}{\Lx},
\end{equation}
which is the probability to have a tracer on a given site ($\Lx^{-1}$) and a bath particle on its right ($\rhoz$), multiplied by the rate ($p$) at which they exchange positions, with an analogous opposite contribution of the tracer exchanging position with a bath particle on its left with a rate $q$.
Finally, the total current can be obtained by a similar argument, or by using the Relation~\eqref{eq:jjbvtr},
\begin{equation}
\label{eq:jeps1}
\cJ = \frac{(p-q) (1-\rhoz)}{\Lx}.
\end{equation}
It can again be interpreted as the probability $\Lx^{-1}$ to have a tracer on a given site, multiplied by the probability $1-\rhoz$ that the site on its right is empty, and  by the rate $p$ at which they exchange positions, with an analogous opposite contribution for the case of an empty site on the left of the tracer.

\subsection{General $\epsilon$}

In this section we write approximate equations to compute the density profile for any $\epsilon$. We start with the evolution equations for the densities,
\begin{widetext}
\begin{eqnarray}
\label{eq:dttau}
\frac{\dd \rho_1}{\dd t} &=& \langle (1+p (1-\tau_1) + \epsilon p \tau_1) \tau_{2} - (1+q (1-\tau_{\Lx-1}) + \epsilon q \tau_{\Lx-1}) \tau_{1} + \epsilon (q \tau_{\Lx-1} - p \tau_1) \rangle \nonumber \\
\frac{\dd \rho_x}{\dd t} &=& \langle (1+p (1-\tau_1) + \epsilon p \tau_1) (\tau_{x+1}-\tau_x) + (1+q (1-\tau_{\Lx-1}) + \epsilon q \tau_{\Lx-1}) (\tau_{x-1}-\tau_x) \rangle \qquad x=2,\ldots,\Lx-2 \nonumber \\
\frac{\dd \rho_{\Lx-1}}{\dd t} &=& \langle - (1+p (1-\tau_1) + \epsilon p \tau_1) \tau_{\Lx-1} + (1+q (1-\tau_{\Lx-1}) + \epsilon q \tau_{\Lx-1}) \tau_{\Lx-2} - \epsilon (q \tau_{\Lx-1} - p \tau_1) \rangle.
\end{eqnarray}
\end{widetext}
Two-point correlations appear on the RHS of equations~\eqref{eq:dttau}. In order to close the equations we make the simplifying assumption that the connected part of the correlations vanishes in the $\Lx\rightarrow\infty$ limit, giving $\langle \tau_x \tau_{x'} \rangle \simeq \rho_x \rho_{x'}$. 

We adapt the definitions of $\Ap$ and $\Am$ to this case, $\Ap = 1+p(1-\rho_1)+\epsilon p \rho_1$ and $\Am = 1+q(1-\rho_{\Lx-1})+\epsilon q \rho_{\Lx-1}$.
After averaging and in the stationary state, writing $h_x = \Ap \rho_{x+1} - \Am \rho_{x}$, Eqs.\,\eqref{eq:dttau} become 
\begin{eqnarray}
\label{eq:rhostat}
h_1 + \epsilon (q \rho_{\Lx-1} - p \rho_1) &=& 0 \nonumber \\
h_{x+1} - h_x &=& 0 \qquad x=2,\ldots,\Lx-2 \nonumber \\
-h_{\Lx-2} - \epsilon (q \rho_{\Lx-1} - p \rho_1) &=& 0.
\end{eqnarray}
The bulk equation is solved by a constant $h_x=C_1$, and the boundary equations both give $C_1 = \epsilon (p \rho_1 - q \rho_{L-1})$, such that a partial solution is
\begin{equation}
\label{eq:rhopart}
\rho_x = \epsilon \frac{p \rho_1 - q \rho_{\Lx-1}}{\Ap-\Am} + \left( \rho_1 - \epsilon \frac{p \rho_1 - q \rho_{\Lx-1}}{\Ap-\Am} \right) \left( \frac{\Am}{\Ap}\right)^{x-1}.
\end{equation}
After replacing $\rho_1$ and $\rho_{\Lx-1}$ by their values in~\eqref{eq:rhopart}, there remain two quantities to be determined, $\Ap$ and $\Am$. The first one is obtained by evaluating~\eqref{eq:rhopart} for $x=\Lx-1$ and replacing $\rho_{\Lx-1}$ by its value in terms of $\Am$ on the LHS. The second one is the normalization $\sum_{x=1}^{\Lx-1} \rho_x = N$. These two equations give
\begin{widetext}
\begin{eqnarray}
\label{eq:apam}
\frac{1+q -\Am}{q} (\Ap-\Am) - \epsilon (p-\Ap -q + \Am) &=& \left( \frac{1+p -\Ap}{p} (\Ap-\Am) - \epsilon (p-\Ap -q + \Am)\right) \left( \frac{\Am}{\Ap}\right)^{\Lx-2}, \\
N (\Ap-\Am) (1-\epsilon) - (\Lx-1) \epsilon (p-\Ap -q + \Am) &=& \left( \Ap \frac{1+p -\Ap}{p} - \epsilon \Ap \frac{p-\Ap-q+\Am}{\Ap-\Am}\right) \left( 1- \left( \frac{\Am}{\Ap}\right)^{\Lx-1}\right), \nonumber
\end{eqnarray}
\end{widetext}
where we can check that $\Ap-\Am = O(\Lx^{-1})$ for $\epsilon = 0$. For $\epsilon = 1$ we recover $\Ap = 1+p$, $\Am = 1+q$.

In general, Eqs.\,\eqref{eq:apam} have to be solved numerically. For large $\Lx$ and large enough $\epsilon$ the system simplifies considerably, as both RHSs are subdominant. 
Equations~\eqref{eq:apam} are then easily solved to yield
\begin{eqnarray}
\label{eq:apamsol}
\Ap &=& 1+q (1-\rhoz) +\epsilon q \rhoz + \epsilon \frac{p-q}{\rhoz + \epsilon(1- \rhoz)},\nonumber \\
\Am &=& 1+q (1-\rhoz) +\epsilon q \rhoz .
\end{eqnarray}
In particular, the second of the Eqs.\,\eqref{eq:apamsol} states that the density at site $\Lx-1$ is simply $\rhoz$, which strongly resembles the behavior obtained in Eq.\,\eqref{eq:decaysumdiscr} for the density of the 2D model projected on the $x$ axis.
  From Eq.\,\eqref{eq:apamLinf} the decay length becomes
\begin{eqnarray}
\label{eq:decaylength1d}
\frac{1}{\log \Ap -\log \Am} &=& \frac{1}{\log \left( 1+ \epsilon \frac{p-q}{(\rhoz + \epsilon(1- \rhoz)) (1+q (1-\rhoz) +\epsilon q \rhoz)} \right)} \nonumber \\&=& \frac{1}{\epsilon} \frac{\rhoz (1+q-q \rhoz)}{p-q} + o(\epsilon^{-1}).
\end{eqnarray}
It diverges at small $\epsilon$ and at small $p-q$ for an infinite system. The density in front of the tracer is given by
\begin{equation}
\label{eq:rho1}
\rho_1 = \frac{A_+ -1 -p}{p(1-\epsilon)} = \frac{\rhoz (p-q(1-\rhoz)(1-\epsilon))}{p(\rhoz+\epsilon(1-\rhoz))}.
\end{equation}

\begin{figure}[!ht]
	\begin{center}
		\includegraphics[width=0.45\textwidth]{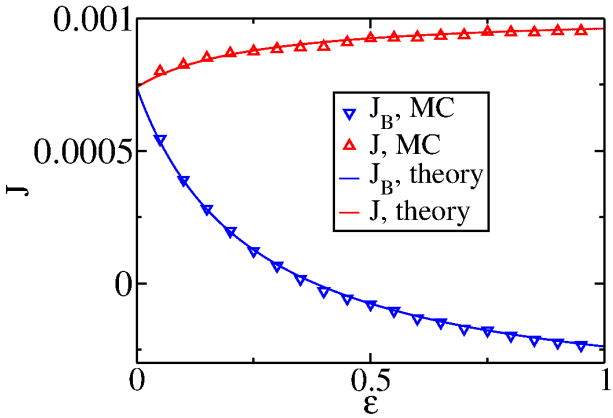} 
	\end{center}
	\caption{\small Bath particle and total currents $\cJb$ and $\cJ$ as a function of $\epsilon$ for $p=1.3$, $q=0.7$, $N=99$ and $\Lx=500$. The theoretical curves are given by Eqs.\,\eqref{eq:jbepsany} and~\eqref{eq:jepsany}.}  
	\label{fig:J}
\end{figure}

\begin{figure}[!ht]
	\begin{center}
		\includegraphics[width=0.45\textwidth]{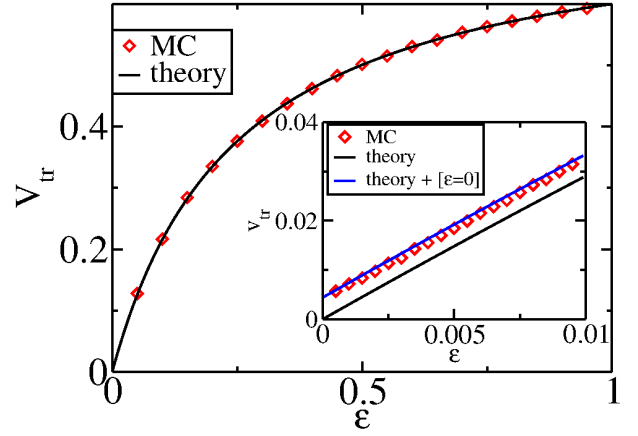} 
	\end{center}
	\caption{\small Velocity of the tracer, $\vtrssep$, as a function of $\epsilon$ for $p=1.3$, $q=0.7$, $N=99$ and $\Lx=500$. The theoretical curve is given by Eqs.\,\eqref{eq:vtrepsany}. Inset: Same for small values of $\epsilon$. The expression~\eqref{eq:vtrepsany} (in black) is compared to~\eqref{eq:vtrepsanyeps0} (blue). }  
	\label{fig:vtr}
\end{figure}

Knowing the density profile, one can also compute the currents. The velocity of the tracer becomes
\begin{equation}
\label{eq:vtrepsany}
\vtrssep = A_+ - A_- = \frac{(p-q) \epsilon}{\rhoz + \epsilon(1-\rhoz)}.
\end{equation}
For the current of bath particles we find
\begin{equation}
\label{eq:jbepsany}
\cJb = \frac{(p-q) \rhoz}{\Lx} \frac{(1-\rhoz) (1-\epsilon)-\epsilon p}{p (\rhoz+\epsilon(1-\rhoz))},
\end{equation}
and the total current is given by
\begin{equation}
\label{eq:jepsany}
\cJ = \frac{(p-q) (1-\rhoz)}{\Lx} \frac{\rhoz (1-\epsilon) +\epsilon p}{p (\rhoz+\epsilon(1-\rhoz))}.
\end{equation}
The expressions~\eqref{eq:jbepsany}, \eqref{eq:jepsany} and~\eqref{eq:vtrepsany} are compared to numerical results in figures~\ref{fig:J} and~\ref{fig:vtr}.
We notice that we get Eqs.\,\eqref{eq:vtreps1}, \eqref{eq:jbeps1} and~\eqref{eq:jeps1} back when taking $\epsilon=1$ in~\eqref{eq:vtrepsany}, \eqref{eq:jbepsany} and~\eqref{eq:jepsany}, respectively. 

However, the small $\epsilon$ regime is not well described (see Fig.\,\ref{fig:vtr}) by the factorization approximation. In particular, the expected velocity~\eqref{eq:vtrepsany} vanishes for $\epsilon=0$, and the predicted value of both currents is $\frac{(p-q)(1-\rhoz)}{\Lx p}$, in contradiction with the exact results~\eqref{eq:vtreps0} and~\eqref{eq:jbjeps0}. In this limit the velocity of the tracer is better described by adding the $\epsilon=0$ contribution~\eqref{eq:vtreps0} to the nonzero $\epsilon$ prediction~\eqref{eq:vtrepsany}. For small $\epsilon = O(N^{-1})$,
\begin{equation}
\label{eq:vtrepsanyeps0}
\vtrssep|_{\epsilon = O(N^{-1})} \simeq  \frac{(p-q) \epsilon}{\rhoz + \epsilon(1-\rhoz)} + \frac{\ust}{N},
\end{equation}
where $\ust$ is still implicitly defined by Equation~\eqref{eq:ustimpl}. The two terms in this equation correspond to the two ways the tracer can move forward, either by exchanging positions with a particle on a site next to it ($\epsilon$ term) or pushing the whole system forward ($N^{-1}$ term). For $\epsilon = O(N^{-1})$ these two contributions become of the same order, and Eq. \eqref{eq:vtrepsany} has to be corrected. 

\section{Hard disks in a two-dimensional narrow channel}
\label{section:harddisks}

After having studied relatively simple lattice systems, we now turn to a more realistic continuum setup, namely hard disks (or 'HD' for short) in a narrow channel, which obey Langevin dynamics. We find a systematic way to make a correspondence between both systems, which enables us to use the results of the previous section~\ref{section:1dssep}.

\subsection{Model definition}

We consider $N+1$ hard disks of diameter $\sigma$ in a narrow channel of length $\Lx$ and periodic in the $X$ direction, and of width $\Lyp = \Ly + \sigma$ with thermal boundaries in the $Y$ direction. 
The positions of their centers and their velocities are denoted by $\bR_k(t) = (X_k(t),Y_k(t))$ and $\bV_k(t) = (V_{k,X}(t),V_{k,Y}(t))$, respectively, where $k$ ranges from $0$ to $N$. The particles are assumed to obey the Langevin equation
\begin{eqnarray}
\label{eq:hddyn}
\frac{\dd\bR_k}{\dd t} &=& \bV_k, \\
\frac{\dd\bV_k}{\dd t}  &=& \bF_k - \gamma \bV_k  + \sqrt{ 2 \gamma k_B T} \bxi_k, \nonumber
\end{eqnarray} 
where $T$ is the temperature, $\gamma$ is the damping coefficient common to all particles, and the $\bxi_k$ are delta-correlated white noises. Only one force does not vanish, say $\bF_0$, and we take it to be parallel to the $X$ direction, $\bF_k = \delta_{k,0} F \bex$. Particle $k= 0$, therefore, is the tracer particle. 
It is clear that changing the values of the diameter $\sigma$ and the temperature $T$ are equivalent to rescaling time and space. In the following we therefore consider only the case $\sigma=1$ and $k_B T = 1$. 

Numerically, the motion equations may be solved to first order in 
the time step $\Delta t$ according to \cite{gillespie1996a, gillespie1996b}
\begin{eqnarray}
\label{eq:hdsimu}
\bR_k(t + \Delta t) & = &  \bR_k(t)  + \bV_k(t) \Delta t, \\
\bV_k(t + \Delta t) & = &  \bV_k(t) + (\bF_k -\gamma \bV_k(t))\Delta t  + (2 \gamma \Delta t)^{1/2} \bn_k, \nonumber
\end{eqnarray}
for all $k=0,\ldots,N$ where $\bn_k$ is a normally-distributed random variable.

The particles are hard disks, which means that all configurations where $|\bR_k-\bR_l| < 1$ for any pair $k$, $l$ of particles are forbidden.
Putting the origin into the center of the simulation box, 
the periodic boundary  in $X$ direction requires that $| X_k | < L_x / 2$. The boundaries in the $Y$ direction only allow particle 
configurations for which $ | Y_k | <  (\Lyp - 1)/2 $. 
These constraints imply that no overtake is possible in channels with $\Lyp < 2$. Collisions between two particles are elastic,
while the walls at $Y= - \Lyp/2$ and $Y= \Lyp/2$ are van Beijeren thermostats ~\cite{vanbeijeren2014}
at a temperature $T = 1$ . Note that thermostatting these walls is not really necessary, since the Langevin dynamics~\eqref{eq:hddyn} is capable to generate nice stationary states without additional thermostatting at the boundary.

This system is studied by \textit{molecular dynamics} (MD) simulations. In the simulations we compute the local one-dimensional density field $\rho(x)$ of neutral particles in a frame co-moving with the tracer projected on the $X$ direction. As in the previous sections, $x$ is the $X$-separation from the tracer, also called \textit{reaction coordinate} in the context of hard disks~\cite{bowles_m_p2004}.  
We also compute  the local pressure 
$P_{yy}(x)$ at the thermostatted boundary at a distance $x$ from the tracer. It corresponds 
to the rate of $y$-momentum transfer per unit wall length
due to the neutral particles colliding with the wall at a distance $x$ from the tracer: 
\begin{eqnarray}
\label{eq:defP}
P_{yy}(x) &=& \frac{1}{2 \delta x \tau} \int_0^{\tau} \dd t \int_{x - (\delta x/2)}^{x+(\delta x/2)} \dd x'  \sum_{k=1}^N \\ 
&& \times \sum_{c_k}  |V'_{k,Y} - V_{k,Y}|  \delta (t - t_{c_k}) \delta( x' - x_{c_k}).\nonumber 
\end{eqnarray}   
Here, $\sum_{c_k} $ denotes the sum  over all wall collisions of the neutral particle $k$ during the simulation time $\tau$, for $k=1,\ldots,N$.
$\delta x$ is a small reaction-coordinate interval centered at $x$ required for the construction of the  histogram, and
$t_{c_k}$ and $x_{c_k}$ are the time and the reaction coordinate of the respective collision of particle $k$ with the wall. Finally, $V'_{k,y} - V_{k,y}$ is the difference between the $y$ component of the velocity of particle $k$ before and after the collision. The factor $2$ in the denominator accounts for the upper and lower boundaries. 

\begin{figure}[th]
	\centering
	{\includegraphics[width=0.42\textwidth]{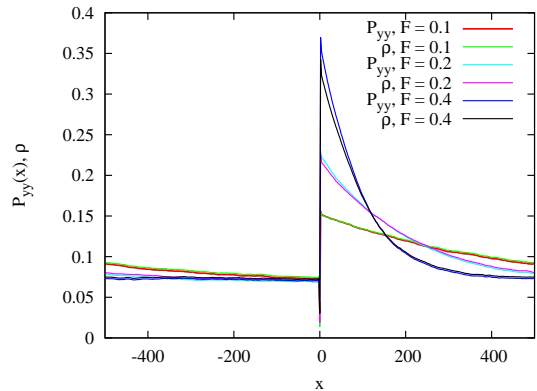}}
	\caption{ Local pressure $P_{yy}$ and local density $\rho$ as a function of the separation $x$ from the tracer
		for various external forces $F = 0.1, 0.2$ and $0.4$.  
		The channel is of  length $L_x = 1000$ and of width  $L_y' = 2.05$.  $999$ neutral particles
		and a single tracer at a temperature $ k_B T = 1$ are considered.  }
	\label{p_rho}
\end{figure}
As an example we compare in Fig. \ref{p_rho} the local pressure  and the local density  as a function of the
distance $x$ from the tracer. The driving force varies between $F =0.1$ and $F  = 0.4$. The field  $F = 0.2$ is small enough such that the ideal-gas equation of state is well obeyed. Therefore, only the local density is considered in the following. examples, and the field is restricted to $0.2$.

The time-averaged velocity in $x$ direction of the tracer is defined by
\begin{equation}
\label{eqref:vtrhd}
\vtrhd = \frac{ X_0(\tau) - X_0(0) } {\tau}.
\end{equation}

In order to compare the results of this model with those of the discrete SSEP approach from the previous section, we have to find a way to apply the results of Section~\ref{section:1dssep} to the hard disks system.  More precisely, we would like to find a mapping between the sets of SSEP parameters $(N,\Lx,p,q,\epsilon)$  and 
HD parameters $(N,\Lx,\Ly,F,\gamma)$.

\subsection{Narrow channel without overtake: The SSEP parameters $p$ and $q$ }

\begin{figure}[th]
	\centering
	{\includegraphics[width=0.4\textwidth]{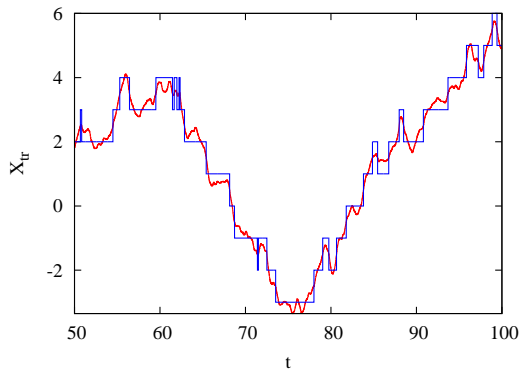}}
	\caption{ Coarse grained motion of the tracer along $X$. The red curve is the actual position of the center of the tracer, while the blue line is the coarse-grained picture of a tracer hopping from site to site. For the counting of the up- and down-steps (see the main text) the periodicity of the boundary in $X$ direction needs to be unfolded. No neutral particle is required.  }
	\label{fig:steps}
\end{figure}

We first note that in the SSEP we want all the particles to have the same 'temperature', which requires to take $\phd+\qhd=2$, where the $\mathrm{HD}$ superscript indicates that that the values of $\phd$ and $\qhd$ are those equivalent to the hard disks system.

The computation of the parameters $\phd$ and $\qhd = 2-\phd $ is straightforward, as they can be defined as the rate at which the tracer hops to the 
right and to the left in the limit of vanishing density. We, therefore, consider only a single tracer without any neutral particle.The simulation box of
length $\Lx$ is partitioned into $\Lx$ boxes of unit length, and the continuous trajectory is replaced by
up-steps and down-steps to the box centers, whenever a particle crosses to a neighboring box 
to the right or the left, respectively. See  Fig. \ref{fig:steps}. We denote  the sum of all up-steps (down-steps) of the  
tracer by $n_{+}^{\mbox{tr}}$ ( $n_{-}^{\mbox{tr}}$ ).  For $F > 0$, $n_{+}^{\mbox{tr}}  >  n_{-}^{\mbox{tr}}$ holds.
The hopping rates for the tracer in ($\phd$) and against ($\qhd$) the field direction become 
\begin{equation} 
\label{eq:mappq}
\phd =  \frac{ 2 n_{+}^{\mbox{tr}}}{ n_{+}^{\mbox{tr}} +n_{-}^{\mbox{tr}} }  \;\; \mbox{respective} \;\;
\qhd =  \frac{ 2 n_{-}^{\mbox{tr}}}{  n_{+}^{\mbox{tr}}  +n_{-}^{\mbox{tr}}},
\end{equation}
where we have used that $\phd + \qhd = 2$. 

When comparing dynamical quantities such as the tracer velocity, one must make sure that the definitions of time are consistent between both systems. In order to do so, we rescale the time in the HD system such that the tracer's total rate of hopping is $\phd + \qhd = 2$. This implies
that the original time $t$ and the rescaled time $t^{HD}$ are related by
\begin{equation}
\label{eq:mapt}
\thd =  t \approx \frac{ n_{+}^{\mbox{tr}} +n_{-}^{\mbox{tr}}}{2}.
\end{equation} 
As a consequence, the rescaled temperature becomes $k_B T^{HD} = 4 k_B T$,

The results for the densities  
studied here are listed in Table~\ref{table:eps0}. This procedure gives values of $\phd-\qhd$ close to what is obtained by equating the average velocities of the free tracers in both cases, $\phd-\qhd \simeq 2 \frac{F}{\gamma}$. 

\begin{table}[htbp]
	\caption{Some parameters and results for $F=0.2$, $\gamma=2$, $\Lyp = 1.90$, $\Lx = 1000$ and varying $N=49, 79, 99$. The SSEP velocity is obtained by using Eq.\,\eqref{eq:vtreps0} with the values $\phd$ and $\qhd$ for the hopping rates and the density $\rho = \frac{N}{\Lx-1}$. The associated density profiles are plotted in Fig.\,\ref{fig:density_N}.}
	\label{table:eps0}
	\begin{tabular}{l|c|c|c} \hline \hline
		$N                $       & 49             &  79                 &  99           \\
		\hline
		$   \vtrhd \times 10^3          $  &  $ 4.16 \pm 0.08 $  &  $   2.590 \pm    0.046  $  & $1.994 \pm 0.044$      \\
		$\phd$                &  1.11255             &  1.11255    &   1.11255               \\
		$\qhd$                 &  0.88745            &    0.88745   &   0.88745                 \\
		$\vtrssep \times 10^3$    &   4.032          &   2.484    &      1.960            \\
	\end{tabular}
\end{table}

\begin{figure}[th]
	\centering
	{\includegraphics[width=0.4\textwidth]{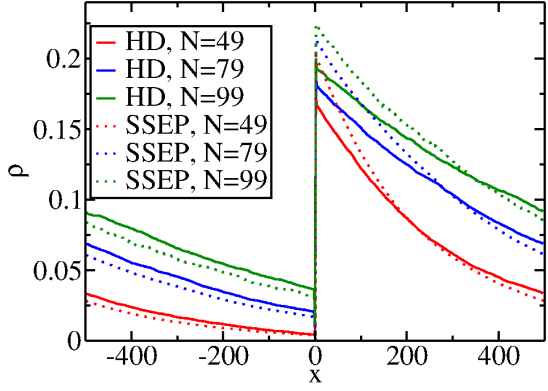}}
	\caption{Local density in the co-moving frame of the tracer for  various $N$. The channel widths  $\Lyp = 1.90$.
		Particles cannot pass each other, and $\epsilon = 0$.
		The driving field  $F = 0.2$. The length of the  periodic channel $\Lx = 1000$. The associated data is found in Table~\ref{table:eps0}.}
	\label{fig:density_N}
\end{figure}

To test this correspondence, we consider channels with a  width $\Lyp < 2 $, which do not allow particles to overtake, and for which 
the SSEP parameter $\epsilon$ vanishes.
In this case the local pressures, densities and the tracer velocity $\vtrhd$ have been checked to be almost independent of $\Lyp$.
Similarly, the dependence of the $p$ and $q$ values obtained from equation~\eqref{eq:mappq} is very weak.

We now fix $\Lyp = 1.9$ and vary the number of particles (and, consequently, also the global density).
Various simulations with different particle numbers $N$ and  a single tracer are carried out.
The particle density profile in the co-moving frame of the tracer is shown in Fig. \ref{fig:density_N}. The tracer velocity $\vtrhd$ is shown in Table~\ref{table:eps0} and shows good agreement with the expression from the SSEP.

\subsection{ The parameter $\epsilonhd$ for channels with overtake}


\begin{figure}[th]
	\centering
	{\includegraphics[width=0.42\textwidth]{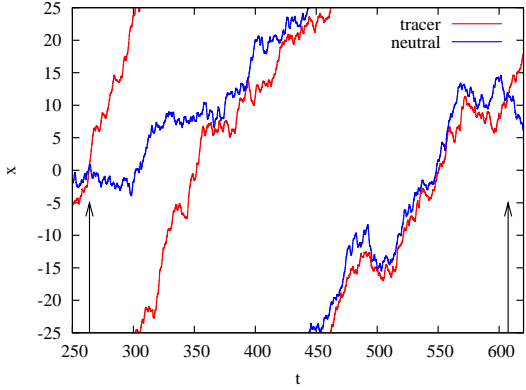}}
	\caption{ Evolution of a tracer particle (red) and of a neutral particle (blue) along the channel in $X$ direction
		as a function of time $t$. The channel width $\Lyp = 2.1$. Thus, overtake events occur at
		times indicated by the black vertical arrows.   Here, the length of the periodic channel ($\Lx$ is $50$) 
		rsnges from $X = -25$ tp $X = 25$.	 }
	\label{fig:traj1}
\end{figure}

We now want to allow the hard disks to overtake, \textit{i.e.} we introduce a nonvanishing parameter $\epsilonhd$ analogous to the SSEP parameter $\epsilon$,
such that $\epsilonhd \phd$ and $\epsilonhd \qhd$ become the overtake rates in and against the field direction, respectively. This is done by widening the channel, 
where we expect the $\epsilonhd$ parameter of the SSEP to be closely related to $\Lyp-2 \sigma$.

\begin{table}[htbp]
	\caption{Some parameters and results for $F=0.2$, $\gamma=2$,  $\Lx = 1000$, $N=99$ and various
		channel widths, $\Lyp= 2.02, 2.05, 2.10$. 
		The case $\Lyp = 1.9$ is contained in Table \ref{table:eps0}.
		The SSEP velocity is obtained by using Eq.\,\eqref{eq:vtrepsanyeps0} with the values $\phd$, $\qhd$ and $\epsilonhd$for the hopping rates and the density $\rho = \frac{N}{\Lx-1}$ taken from the HD simulations. The associated density profiles are plotted in Fig.\,\ref{fig:witheps}.
	}
	\label{table:witheps}
	\begin{tabular}{l|c|c|c|c} \hline \hline
		$\Lyp                $        &  2.02                 &  2.05  & 2.10         \\
		\hline
		$   \vtrhd  \times 10^3   $ &  $2.86 \pm 0.04  $&  $6.64 \pm 0.1  $& $18.06  \pm 0.16$    \\
		$\phd$                &  1.114    &   1.115   &     1.114       \\
		$\qhd$                 &    0.886   &   0.885   &    0.886          \\
		$\epsilonhd$                 &    0.00037   &   0.00219   &     0.0081         \\
		$\vtrssep  \times 10^3$    &   2.87   &    7.01   &   19.54       \\
	\end{tabular}
\end{table}

\begin{figure}[th]
	\centering
	{\includegraphics[width=0.42\textwidth]{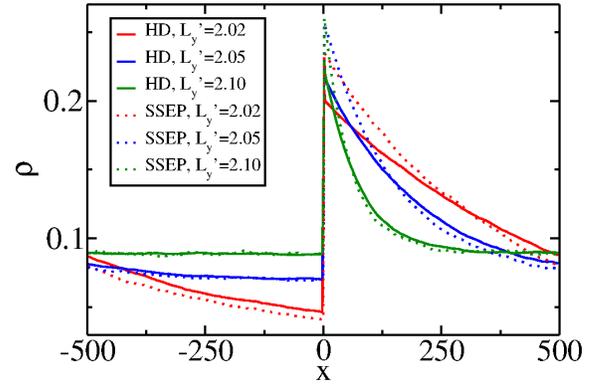}}
	\caption{Local density in the co-moving frame of the tracer for the same systems as in Table~\ref{table:witheps}, compared to their SSEP equivalents given in 
		the table.}
	\label{fig:witheps}
\end{figure} 

For the estimation of $\epsilonhd$ in the limit of low density, a simulation with only two particles
(the tracer and a neutral particle) suffices.
In Fig.~\ref{fig:traj1} the $X$ coordinates of the two particles are shown as a function of time.  The simulation 
box is periodic in $X$ and of length $L_x = 50$ (for demonstration purposes only).
We call an overtake event ''regular'' (r), if the tracer approaches the neutral  particle in the direction 
of the field (from below in Fig. \ref{fig:traj1}, and irregular (i) otherwise. As expected, regular overtakes dominate.  One observes in
Fig.~\ref{fig:traj1} that the tracer (red) collides with the neutral particle quite often from below and  pushes it in the 
field direction, before a regular overtake occurs. The times of two consecutive regular  overtake events  are 
indicated by the black vertical arrows. 

In the following we consider all time intervals between consecutive overtake events.
The initiating and finishing overtakes of each interval may be of regular $(r)$  or irregular $(i)$ type. We characterize such an
interval by the indices $\alpha, \beta$, where  the first index $\alpha \in \{r,i\}$ refers to the type of the initiating,  the
second index  $\beta \in \{r,i\}$ to that of the  closing overtake of that interval. 
The total number of intervals of type  $\alpha, \beta$ is denoted by $o_{\alpha\beta}$, and the {\em average} number of
hard-core collisions of the tracer with the neutral disk during an interval of that type, which will not result in overtake, is called $n_{\alpha \beta} $. The total 
number of collisions in all $\alpha,\beta$-type intervals is  $N_{\alpha \beta} = o_{\alpha \beta} n_{\alpha \beta} $. 
For the long channels considered here, we find that $n_{ir} = n_{ri}$,  and $n_{ii} = 0$. Thus, no successive irregular
overtakes occur.   Since $\epsilonhd \phd $ is the rate with which the
tracer overtakes a neutral particle on its right in the direction of the field, we are only concerned 
with the intervals of type $r,r$ and $i,r$,  which are terminated by regular overtakes.  Similarly, the opposite case of the rate of irregular tracer overtakes
against the field,  $\epsilonhd \qhd$, only involves intervals of type $r,i$ with an irregular ending.
Therefore, there are two ways to estimate  $\epsilonhd$, 
\begin{equation}
	\phd  \epsilonhd = \frac{o_{rr}+o_{ir}}{N_{rr} + N_{ir}}\;\;\mbox{and}\;\; \qhd  \epsilonhd = \frac{o_{ri}}{N_{ri} }.
	\label{eps}
\end{equation}
Since $\phd + \qhd = 2$, they may be combined to give
\begin{equation}
	2 \; \epsilonhd = \frac{o_{rr}+o_{ir}}{N_{rr} + N_{ir}} +  \frac{o_{ri}}{N_{ri} }.
	\label{eps_final}
\end{equation}
The results for $\epsilonhd$ from Eq. (\ref{eps_final}) are listed in Table \ref{table:witheps}.
For small $\Lyp$ the SSEP velocity $\vtrssep $ computed with the parameters  $\phd, \qhd$ and $\epsilonhd$
(which are also given in Table \ref{table:witheps}) agrees very well with the time-averaged velocity $   \vtrhd $ from the HD simulations.
However, for larger channel widths this agreement becomes gradually worse. The associated case for vanishing $\epsilonhd$ is 
included in Table \ref{table:eps0}.

A fit of the values for $\epsilonhd$ listed in Table~\ref{table:witheps} gives a quadratic dependence on the width, $\epsilonhd \sim 0.814 (\Lyp-2)^2$. This is compatible with simple geometric arguments. Density profiles and tracer velocities in the HD system and in the SSEP equivalent are compared in  Fig.\,\ref{fig:witheps}.
It is, perhaps,  surprising that the agreement for the density profiles  becomes worse for very narrow channels, whereas the opposite is true for the mean velocities. 
We do not yet know the reason for this behavior. 

\section{Summary}
\label{section:sum}

In this paper we analyzed the steady state properties of a driven tracer moving in a
two dimensional narrow channel, by using a combination of lattice gas models and of a continuum model
of hard disks. The force-velocity relation has been calculated as well as the density and pressure profiles
in the channel. Considerable non-homogeneous density and pressure profiles at the boundaries of narrow
channels have been observed.

Three models have been studied. The first model, the 2D SSEP, is simple enough to give analytical information about the bidimensional spatial structure of the density disturbance created by the tracer. In the frame of the tracer a simple picture emerges, with the tracer behaving like a dipolar source. Its long-range effects are however screened by two mechanisms. The first is the flow of the surrounding fluid which yields a $3/2$ power law decay of the density disturbance at the back of the tracer and exponential decay in all other directions. The algebraic decay is further screened by the width of the channel $\Ly$, thus leading to exponential density profile at distances larger than $\Ly$. The 2D SSEP however has the drawback that the width of the channel is discrete and does not really allow to understand the small width regime.
We therefore turned to analyze a  second model, a 1D SSEP with overtakes allowed (which simulate available paths around the tracer in a narrow channel). In this model the density and pressure profiles, the currents and the velocity of the tracer can be calculated with good accuracy. 

In order to check the validity of the lattice gas models, we also performed molecular dynamics simulations for a continuous system of overdamped hard disks in a narrow channel. We found that the behavior of this system is  very similar to that of the lattice gases.
Making a correspondence between  the parameters of the lattice gas models to those of  the hard disk gas, a good agreement for the  density profiles, pressure profiles and tracer currents has been obtained.

This work suggests some extensions. In the 2D case, a natural extension would be to introduce a second tracer particle and to study interactions between the tracers. While the discrete problem seems hard to tackle, simplified continuous equations like~\eqref{eq:rhoc} may reproduce the main features of a two- or even many-tracer system. One may also generalize the molecular dynamics to consider disks with significant inertia and to study  how they correspond to lattice gas models.

\section*{Acknowledgments}
We thank O. Hirschberg, Y. Kafri and A. Kundu for useful discussions. One of us (HAP) wants to acknowledge the hospitality and support of the Weizmann Institute of Science, where parts of the work reported 
here have been performed. The support of the Israel Science Foundation (ISF)
is gratefully acknowledged. The molecular dynamics simulations were carried out on the Vienna Scientific Cluster (VSC1, VSC2, and VSC3).
We are grateful for the generous allocation of computer resources.

%

\end{document}